# SPIN POLARONS AND HIGH-$T_c$ SUPERCONDUCTIVITY

**A. L. Chernyshev**[†,*] **and R. F. Wood**
Solid State Division, Oak Ridge National Laboratory, P.O. Box 2008, Oak Ridge, Tennessee 37831

## 1. INTRODUCTION

A charge carrier (electron or hole) moving through an ionic lattice is always accompanied by displacements of the ions. Under certain conditions, the carrier plus these displacements can form a good quasi-particle, i.e., the ionic polaron. In an analogous manner, a spin polaron is a charge carrier moving in a magnetic medium accompanied by deviations of localized spins. Before the era of high-$T_c$ superconductivity (HTSC), the research effort on spin polarons was quite modest. Now with HTSC in the cuprates and colossal magnetoresistance in the manganites, this area is experiencing an explosive growth. Further rapid expansion is expected as other complex oxides, magnetic semiconductors, and related materials are studied.

High-temperature superconductivity in the cuprates was discovered just over fifteen years ago [1-3], but theoretical understanding of its origin is still far from complete. Nevertheless, the theoretical effort to understand the fundamental mechanism has already led to the development of an impressive array of intricate concepts and sophisticated mathematical techniques. The notion of the spin polaron has played an important role in many of these developments. In this article, we consider aspects of the spin-polaron concept as they relate to HTSC. The article is intended to be an overview in which a flavor of the research on spin polarons is given. The emphasis will be on the intuitive physical picture; for the more detailed mathematical formalism, references to the original works or to other review articles will be given.

Choosing stoichiometric $La_2CuO_4$ as the prototypical high-$T_c$ material, the conceptual framework on which the article is based can be summarized as follows.

La$_2$CuO$_4$ is an antiferromagnetic (AF) insulator with the Neel temperature $T_N$=325K and charge-transfer gap $\Delta \cong 2.0\,eV$. The most prominent physical feature of the material is the CuO$_2$ planes, which are "active" electronically and magnetically, and are separated by the rather inert LaO layers. There is only a weak interaction between the CuO$_2$ planes so that the electronic system is essentially two dimensional. Copper ions in the planes form a square lattice with the Cu$^{2+}$ spins interacting antiferromagnetically; the nearest-neighbor exchange interaction is $J \approx 130\,meV$. Thus, La$_2$CuO$_4$ can be described as a layered Mott-Hubbard/charge-transfer insulator [4] with the Cu $3d^{\,9}$ configuration playing a key role and with strong hybridization between the Cu $3d(x^2-y^2)$ lower Hubbard band and the O $2p\sigma$ orbitals. Doping with Sr or Ba introduces holes, predominantly in the O sublattice of the CuO$_2$ planes, which quickly destroys the long-range AF order (2% Sr) and gives rise to the hole-type conductivity and, eventually, to the superconductivity (5%-25% Sr).

The rough physical picture of such an evolution, although not established in all the details, can be understood as follows. At low doping, holes doped into the CuO$_2$ planes form the spin polarons. Local as well as long-range distortions of spins in the antiferromagnetic background are induced. As a consequence, spin-polarons destroy the long-range AF ordering, but short-range correlations persist, as revealed by two-magnon Raman [5] and neutron-scattering experiments [6]. Further doping drives the system through the spin-glass, low-conductivity phase, and then into the superconducting phase with its unusual metallic properties. There are several scenarios of how these properties may develop even within the spin-polaron approach, some of which will be discussed here. We also want to emphasize a more recent development in this area concerning the presence of charge and spin inhomogeneities, also called "stripes", in much of the phase diagram of several HTSC materials. On the theoretical level, it is presently understood that as the hole concentration is increased, a peculiar phase separation into quasi-one-dimensional hole-rich stripes and antiferromagnetic hole-poor regions occurs [7]. The phase of the AF order is shifted by 180º in neighboring hole-poor domains, and thus the metallic stripes are formed at the antiphase domain walls. We will demonstrate how such a behavior can occur as the result of the tendency to form spin polarons in the strongly correlated system of holes and spins.

The plan of the article is as follows. In the next section, a brief introduction to the spin polaron concept and to some aspects of physics of the strongly correlated systems within the "basic" Hubbard and t-J models is given. Two specific realizations of spin polarons, namely ferrons and string polarons, are reviewed. We also briefly consider several other contributions related to superconductivity in the cuprates. We then begin a more in-depth discussion of two of these.

In Section 3, the extended Hubbard model in what might be called a "semiconductor" formulation is considered. Such an approach explicitly imposes a commensurate antiferromagnetic order on the parent material and then calculates the band structure with a Coulomb repulsion (Hubbard $U$) term on the Cu sites included. The spin polaron formation is considered within this context. The gap symmetry and transition temperature as a function of doping concentration are explicitly evaluated.

In Section 4, the body of microscopic studies based on the t-J model is considered. Both numerical and analytical formulations are surveyed. The role of spin polarons in HTSC has been most extensively studied within the framework of the t-J model and this is reflected in the extent of the literature citations in this section. Also, the rapidly growing theoretical literature on stripe formation most often employs the t-J model, frequently with the explicit introduction of spin polarons in both analytical and numerical studies, and this too will be reviewed.

These two sections address much of what is now believed to be the very complex but "essential physics" of the high-$T_c$ materials, presented here in a rather simplified physical picture. By "essential physics" we mean the important physical characteristics of the HTSC systems or their models which are believed to be responsible for the majority of their anomalous properties including the high-$T_c$ phenomenon itself. There is considerable overlap between some of the discussions in Secs. 3 and 4. We thought it useful to retain this overlap as it frequently reflects somewhat different, but not incompatible, approaches to the same problem.

In Section 5, we give a brief summary and a few concluding remarks on the likely direction of research on spin polarons in the near future. Several good review articles on the theory of high temperature superconductivity have appeared [8-11] and some of these will be referred to often below.

## 2. SPIN POLARONS IN STRONGLY CORRELATED SYSTEMS

### a.) Hubbard and t-J Hamiltonians

In this subsection, some of the strongly correlated Hamiltonians used in HTSC and the phenomenology associated with them are reviewed.

As mentioned in Sec. 1, in the conventional band picture the reference parental material of the cuprates, $La_2CuO_4$, has a half-filled conduction band, and hence should be a metal. In reality it is a Mott insulator in which the energy gap for the charge excitations originates from strong electronic correlations. Such a behavior, incompatible with that of simple band insulators, has been studied since the early 1960s in the context of the narrow-band transition metals and their oxides. Models to describe the competition between the correlations and delocalization were proposed by Hubbard [12], Gutzwiller [13], and Kanamori [14]. Of these, the Hubbard model has become the most thoroughly developed. Its Hamiltonian is given by:

$$H = -t \sum_{\langle ij \rangle \sigma} \left( c_{i\sigma}^\dagger c_{j\sigma} + H.c. \right) + U \sum_i n_{i\uparrow} n_{i\downarrow} \quad , \tag{1}$$

where $U$ is the on-site Coulomb repulsion, and $n_{i\sigma} = c_{i\sigma}^\dagger c_{i\sigma}$ is the density operator for electrons at the $i$-th site. There is an extensive literature on studies of the one-band

Hubbard Hamiltonian. It has been reviewed by Herring [15] in an early volume and more recently by Izyumov [16]. In fact, the Hamiltonian of Eq. (1) has become the standard starting point for many mathematical studies of highly correlated electronic systems. One of Hubbard's central results was that in a crystalline array of one-electron atoms with non-degenerate orbitals, the on-site Coulomb repulsion term splits the energy bands into what are now frequently called the upper and lower Hubbard bands and causes the insulating behavior at half-filling. In the beginning of the high-$T_c$ era, Anderson suggested [17] that a doped Mott insulator described by the one-band Hubbard model in the limit of large $U$ should contain the essential physics of the cuprates.

In the strong coupling limit $U >> t$, the system is said to be highly correlated with $U$ associated with the Cu $d$-orbitals. In that case, as shown in Refs. [18,19], by applying a canonical transformation [20] to Eq. (1), the high-energy doubly occupied states can be projected out and one obtains the t-J model,

$$H = -t \sum_{\langle ij \rangle \sigma} \left( P_i^\sigma c_{i\sigma}^\dagger c_{j\sigma} P_j^\sigma + H.c. \right) + J \sum_{\langle ij \rangle} \left( \mathbf{S}_i \cdot \mathbf{S}_j - \frac{1}{4} n_i n_j \right) \ , \qquad (2)$$

where $J=4t^2/U$ is the nearest-neighbor exchange integral, $n_i = n_{i\uparrow} + n_{i\downarrow}$ is the density operator at the site $i$, $\mathbf{S}_i = c_{i\alpha}^\dagger \boldsymbol{\sigma}_{\alpha\beta} c_{j\beta} / 2$ is the spin-density operator, and $\boldsymbol{\sigma} = (\sigma^x, \sigma^y, \sigma^z)$ are the Pauli matrices. We note here that the most important difference of the t-J model from any free-electron model is in the restriction of the Hilbert space. No states with double occupancy of the sites is allowed, which is formally expressed through the projection operators $P_i^\sigma = 1 - n_{i,-\sigma}$ in the kinetic energy term of the t-J Hamiltonian. The t-J model is considered by many as a basic model for the description of the low-energy physics of high-$T_c$ materials. The spin-polaron concept within the t-J model is discussed in Sec. 4.

As noted in Sec. 1, the two-dimensional $CuO_2$ planes represent the essential ingredient of high-$T_c$ superconductors. Also, the role of the O $2p_\sigma$ orbitals is known to be important since the hole doping goes primarily in them [21]. Therefore, an extended (or multi-band) Hubbard model, proposed by several authors [22-24], would seem to be more appropriate for the description of the physical properties of the cuprates. The Hamiltonian for such a model can be written as

$$H = -\sum_{ij,\sigma} \varepsilon_{ij} \left( c_{i\sigma}^\dagger c_{j\sigma} + H.c. \right) + \sum_{ij,\sigma\sigma'} U_{ij} n_{i\sigma} n_{j\sigma'} \ , \qquad (3)$$

in which the diagonal contributions to the first term give the "atomic" energies while the off-diagonal contributions are the hopping integrals. Note that $i$ and $j$ now run over Cu and O sublattices, also there are both diagonal and off-diagonal $U$ terms in the Hamiltonian. Using band calculations in the cuprates as a guide [25], one might conclude that at the very least the Cu $3d(x^2-y^2)$ and the O $2p_x$, $2p_y$ orbitals should be retained, making it a three-band model. This results in a complex Hamiltonian and there have been numerous efforts to show that the two-, three-, and higher-band models can be transformed into the one-band model without losing the essential physics at low energies [26-31]. In these approaches, the O orbitals are "folded" onto nominal Cu-sites and disappear explicitly from an effective Hamiltonian. The coupling between the Cu ions goes by way of superexchange, leading to an effective $J$. When additional holes are introduced by doping the holes move predominately on the O ions rather than on the Cu ions. However, the holes on the O and Cu ions can be closely associated to form an entity that moves through the lattice as a singlet in the t-J model, in a manner first discussed by Zhang and Rice [26]. Recently, such a picture has received strong support from spin-polarized photoemission experiments [32]. In order to describe the optical properties of the cuprates, such as the charge-transfer excitations or the ARPES data for the *1-10 eV* energy range [33], one should retain the full orbital information contained in Eq. (3). Spin-polaron effects within the extended Hubbard model are discussed in Sec. 3.

### b.) Spin-polaron concept

The origin of the spin-polaron concept can be traced back to Zener's early work [34] on the "double-exchange" mechanism in manganese oxides, in which doping introduces an effective exchange interaction and results in two neighboring Mn spins being aligned. Double-exchange has been subsequently discussed by Anderson and Hasegawa [35] and especially by de Gennes [36] who demonstrated that the mobile carrier can lead to a polarization of the localized antiferromagnetic spins. Consideration of the single-band Hubbard model with the electronic concentration close to half-filling (one electron per site) has led Nagaoka [37] to the conclusion that this system will have a ferromagnetic ground state in the limit of infinite electron-electron repulsion. While these studies concern different models and seem to be quite different in details, the physical mechanism associated with the net spin polarization is very similar. The kinetic energy of the carrier is optimal when the local spins are aligned. Therefore, if the kinetic energy dominates, it is likely that the charge carrier will form a ferromagnetic "bubble", also referred to as a "ferron" [38]. In the Nagaoka limit the size of the ferron equals the size of the system and thus all the spins are polarized.

We would like to stress here again that both the double-exchange and the Hubbard models are now recognized by many as the "basic" models for a variety of the systems with strong electronic correlations. That is, it is believed that the essential physics of such systems has some universal features described by these models. In particular, both the Hubbard model and the t-J model are thought to be highly relevant to the low-energy physics of the cuprate high-$T_c$ superconductors.

Since the electronic structure of the stoichiometric cuprates corresponds to half-filling with AF ordered local spins, Nagaoka-type spin polarons have attracted much attention in the earlier days of HTSC [39-43]. It turns out, however, that a different type of spin-polaronic solution is possible in the framework of the Hubbard or t-J models, which is more adequate for the actual range of parameters in the cuprates [18,44]. Instead of creating a fully polarized region of spins, which cost energy proportional to the volume of the "bubble" (and to the strength of the spin-spin interaction *J*), the charge carrier may be constrained to oscillate around its origin in a string-like, "retraceable-path" motion [18]. In the string-like motion a disruption of the perfect AF order occurs only along the paths of the charge carrier rather than throughout a "bubble". Therefore, the energy of the string is proportional to its length. This is more energetically favorable in the case where an effective spin-spin interaction is not too small in comparison with the "bare" kinetic energy of the charge carrier. In the t-J model, the comparison would be between the superexchange constant *J* and the hopping integral *t*. Numerical studies as well as analytical considerations, a simplified version of which is given below, show that the Nagaoka-type polarons are favorable when *t/J > 50-200*, while for the realistic systems with *t/J~3* the string-like spin-polaron language seems to be more appropriate. Also, part of the string can be "healed" by the spin fluctuations present in the AF background and such a spin polaron can propagate.

This brings us to a discussion of different kinds of spin polarons. There are two types of distinctions we would like to make. First, is the distinction between the case where an itinerant carrier and a localized spin are in the same orbital state and the case where they are in different orbital states. The single-band Hubbard model and the t-J model are examples of the former case. An example of the latter case is the *s-d* model, where the carrier is in an extended *s*-type orbital with the local spins in *d* orbitals. In this case, the formation of the "ferrons" is likely, as we discuss below. The second distinction is the value of the spin of the spin-polaron. In the case of spin-polarons, the spin of the charge carrier is a combined effect of the spin of the electron (hole) and the polarization carried by the "cloud". For a well-defined "ferron" the spin is large, *S>>1*. In the case of the retraceable-path polaron, no extra magnetic moment is carried by the excitation, therefore *S=1/2*.

Many of the physical systems of interest belong to the class of so-called charge-transfer materials [4,25] where localized magnetic states are associated with the orbitals of the transition metal ions (e.g., *d*-orbitals of $Cu^{2+}$ in cuprates), while the charge carriers are occupying predominantly the orbitals of anions (such as *p*-orbitals of $O^{2-}$). One may suspect then that the "same orbital state" polaron is a rather artificial construction and that, in order to describe realistic materials, one needs to consider a model with all the necessary orbitals involved. However, as it was shown first by Zhang and Rice [26], the two-orbital model for the cuprates can be "mapped" rigorously onto the single-band t-J model. Such a "mapping" implies the projection out of the "high-energy" states to obtain an effective, much simpler model which corresponds to the original model at low energies. Subsequent studies [27-31] have shown that, in fact, such a mapping is possible for a wide class of the models, including the general type three-band Hubbard model proposed for the cuprates [22-24]. It was shown that the result of the reduction is rather universal, that is, the low-energy sector of the many-band Hubbard model coincides with that of the generalized t-J-like model. This "school of thought" has demonstrated the important role of the strong Cu-O hybridization in such a reduction and has ruled out the initial criticism of the Zhang-Rice solution [45]. Roughly speaking, even if one starts with the multi-orbital problem, the strong hybridization between the orbitals can make it identical to the single-orbital case in the low-energy sector of the Hilbert space. Thus, the use of the "same orbital" polaronic ideas is justified for the real systems.

### c.) Examples of the spin-polaronic states

Here we consider in some detail the qualitative features of the Nagaoka-like and string-like spin polarons.

The formation of the Nagaoka polaron is readily understood by considering the simple spin array shown in Fig. 1. It may illustrate the t-J as well as the *s-d* model. For the *s-d* model, the extra electron is assumed to be in an *s* orbital and the localized spins in *d* orbitals. An extra electron with an up spin is introduced into an antiferromagnetically aligned array. If the energy to flip an ↑-spin to a ↓-spin orientation is small, the electron may lower its kinetic energy enough to compensate for the spin-flip energy. Consequently, within some region, the spins become ferromagnetically aligned with one another and antiferromagnetically aligned with the spin of the added electron to form the spin polaron. Under some conditions the extra electron may ferromagnetically align many of the *d* spins, as we have seen for the Nagaoka limit [37].

The formation energy for such spin polarons and their mobility can be estimated in a manner somewhat similar to that of ionic polarons. Since we are concerned here with the properties of the two-dimensional systems, we provide these estimations for the 2D square lattice case. The interaction that produces spin alignment is a kinetic energy effect and it is not directly related to any particular mechanism of the spin-spin exchange. We, therefore, will consider the problem in the framework of the t-J model. We also note that most of the results of Sec. 2(b) and Sec. 2(c) remain valid for other models as long as the physical mechanisms discussed here are applicable.

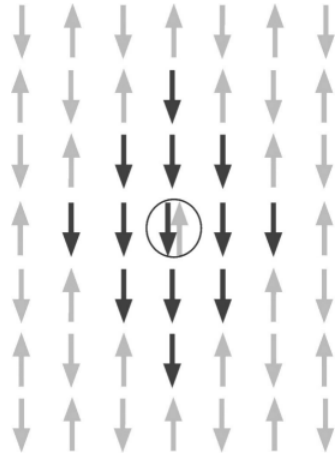

Figure 1. Schematic view of the ferromagnetic region surrounding the extra charge carrier in an AF environment.

The kinetic energy associated with the confinement of the carrier to a 2D disk-shaped region of radius $R$ is:

$$E_{kin} \cong -zt + \frac{\pi^2 b^2}{2m_b R^2} , \qquad (4)$$

where $t$ is a nearest-neighbor hopping integral, $z=4$ is the number of nearest neighbors, $m_b=1/2t$ is the band effective mass of the charge carrier, $R$ is the radius of the ferromagnetic "bubble" (in units of the lattice spacing), and $\beta \approx 1$ is a numerical factor. One half of these spins must be reversed at a cost in energy of $2J$ per reversed spin. The total energy of the spin polaron as a function of $R$ is given by:

$$E = -4t + \frac{t\pi^2}{R^2} + \pi R^2 J . \qquad (5)$$

Minimizing $E(R)$ with respect to $R$, gives the radius of the spin polaron, $R_p$, namely,

$$R_p \cong \left( \pi \frac{t}{J} \right)^{1/4} , \qquad (6)$$

and the polaron energy:

$$E_p = -4t + 2\pi^{3/2} (tJ)^{1/2} . \qquad (7)$$

The total spin of such a polaron is:

$$S_p \cong \frac{\pi R_p^2}{2} = \frac{\pi^{3/2}}{2} \sqrt{\frac{t}{J}} . \qquad (8)$$

A rough estimate of the effective mass of the spin polaron, $m^*_p$, can be obtained from an equation given by Mott and Davis [46], i.e., $m^*_p = m_b \exp(\gamma R_p)$ in which $\gamma \cong 1$ and $m_b$ is a band effective mass. This will be discussed further in the next section.

Another, quite different, way of treating the added charge carrier in the array of spins was introduced by Bulaevskii *et al*. [18]. It is assumed that the added spin forms a tightly bound singlet state with the spin already there. This creates a hole in the spin-*1/2* Heisenberg plane, as shown in Fig. 2. This hole can move without flipping a spin because either of the singlet spins can pair with the nearest neighbor spin as needed, depending on the orientation of that neighbor. However, a trace of spins misaligned relative to the initial AF ordering, a "string", is left behind the hole as indicated in Fig. 2. The energy of such a string is proportional to its length, $E_{string} \propto JL$. Therefore, the hole will be effectively confined in the 1D string potential.

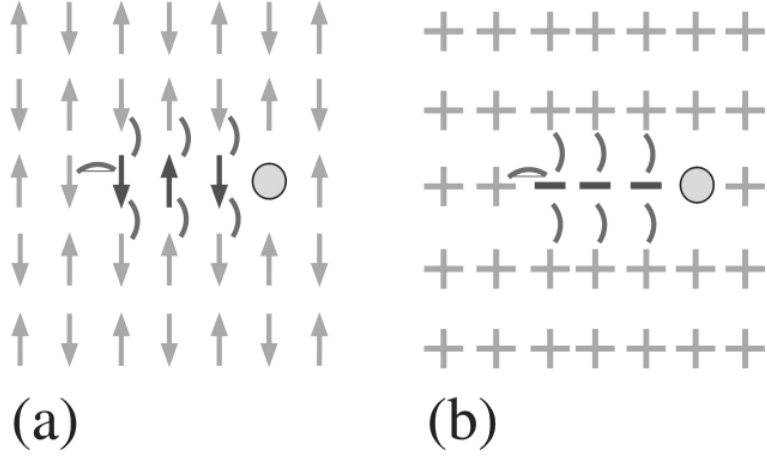

Figure 2. (a) A hole followed by the "string" of misaligned spins in a homogeneous AF environment. (b) same as (a), "+" and "-" denote the sign of the staggered magnetization $M_i=(-1)^i S_i^z$. Arcs denote "wrong" (ferromagnetic) bonds.

It can be shown [18] that in the continuum limit at $t/J \gg 1$, the string spin-polaron problem becomes equivalent to the problem of a single particle in a linear potential with the Hamiltonian:

$$H = -2\sqrt{3}\,t - \sqrt{3}\,t\frac{\partial^2}{\partial L^2} + JL, \qquad (9)$$

where $\sqrt{3}$ originates from the similarity of the hole motion to the motion in the Bethe lattice with the $z-1=3$ coordination number. The solution of the Schrödinger equation for the particle in a linear potential is given in terms of the Airy function. This gives the polaron energy

$$E_p = -2\sqrt{3}\,t + 3^{1/6}\beta_0\, t^{1/3} J^{2/3}, \qquad (10)$$

where $\beta_0 = 2.34$ is the first zero of the Airy function. The average length of the string is $L_p \approx (t/J)^{1/3}$. In considering the string polaron we neglected the presence of the spin flips in the AF background of spins. It is a valid approximation if $t/J \gg 1$ since the hole motion is fast and the spin relaxation is slow in this limit. Therefore, on the fast timescale the hole moves in the background of essentially static, staggered spins. However, such spin flips are important since they provide a mechanism for the spin-polaron delocalization. The spin flips in an AF occur in pairs, that is the spins at two neighboring sites can simultaneously flip. If such a flip happens within the "string", part of the string will be "healed" and the hole is effectively moved. It can be shown that this provides a coherent band for the string-like spin polaron with the bandwidth $W \approx 2J$ [47]. Since $J$ is the subleading order to Eq. (10), this again confirms that the above expression for $E_p$ is correct to the leading order and should be taken as an energy of the bottom of a narrow spin-polaron band.

One can now compare the energies of the spin polaronic states, Eq. (7) and Eq. (10), and estimate the parameter range favorable for Nagaoka or string-like polarons. The numerical solution shows that the Nagaoka polaron is lower in energy if $t/J$ is larger than some critical value $(t/J)_c \sim 350$ [48]. If $t/J < (t/J)_c$, the string polaron is a more favorable excitation. This value of $(t/J)_c$ agrees within an order of magnitude with the more accurate numerical calculation [49] which provides $(t/J)_c \approx 50$.

### d.) Different Approaches to HTSC

Many theoretical approaches to the HTSC problem have emphasized the importance of the experimentally observed magnetism in the parent materials and of magnetic correlations within the superconducting state. In fact, many of the studies have adopted, in one way or another, the AF spin fluctuations as an indispensable ingredient of the theory.

Although the approaches to the problem vary from the semi-phenomenological, experiment-oriented studies [10] to microscopic investigations of strongly-correlated models using sophisticated analytical [50] or powerful numerical techniques [8], their conclusions do not necessarily contradict each other. Rather, they often seem to represent different views on essentially the same properties. One example of such a "unification" is the prevailing theoretical conclusion about the $d(x^2-y^2)$-symmetry of the superconducting order parameter obtained within approaches which involve spin-fluctuation mediated pairing [8-10].

Among these efforts, the spin-polaronic approach has played a prominent role in understanding the basic effects of interplay between the spin and charge degrees of freedom in $CuO_2$ planes [8,51,52]. It is also conceptually very similar to many other theories with respect to the spin-wave mediated pairing mechanism. Although it is generally hard to extract quantitatively valid predictions of experimental results from a microscopic theory, the spin-polaron studies have been successful in doing that. From a purely theoretical viewpoint, this approach also provides a consistent and elegant way of treating many microscopic features of the problem [8,9,53]. Apparently, most of the results of the numerical approaches to the strongly correlated models are to be understood within the spin-polaron paradigm [54-57].

We would like to note here that the spin-polaron approach belongs to the "conservative" group of theories of unconventional superconductors. That is, the excitations are "normal" quasiparticles and the interaction is mediated by a collective mode. Over the last 15 years a number of "unorthodox" approaches to the HTSC problem have been developed, some of which are reviewed in the other chapters of this volume. These include the idea of spin and charge separation of RVB type [58], spin-charge separation within an array of 1D stripes [59], underlying $Z_2$ topological order [50], *d*-density-wave order [60], *SO(5)* symmetry unifying superconductivity and magnetism [61], and others. The spin-polaronic concept is not necessarily orthogonal to these ideas. In fact, as we discuss in Sec. 4, the stripes can be considered as a natural outcome of the same spin-polaronic tendencies in the spin-hole system [57]. Also, additional order parameters, like the *d*-density-wave orbital currents, may come on the top of well-developed spin-polaronic correlations [62].

To summarize this Section, the spin-polaron approach is a viable concept which allows us to explain many aspects of HTSC in a consistent way. Although the theory of high-$T_c$ is yet to be completely established, spin polarons are likely to play an important role in its final form. The next two sections are devoted to more specific details of the spin-polaron approach.

### 3. EXTENDED HUBBARD MODEL

In the extended Hubbard model of Eq. (3), no attempt is made to project out the oxygen orbitals and thus derive an effective Hamiltonian with only Cu-Cu interactions, as is done in the t-J model. Even within the t-J framework it is necessary to relate the effective electronic parameters to those of some model that does include the oxygen orbitals, typically the parameters are taken from a band calculation. Consequently, it is useful to consider the $CuO_2$ planes in more detail than we have done up to this point.

#### a.) Electronic structure of undoped $CuO_2$ planes in the parent materials

An extensive literature on both first principles and parameterized electronic structure calculations for the cuprates developed shortly after high-$T_c$ superconductivity was discovered. The earliest band calculations [63,64] were carried out using established approaches such as the linear augmented plane wave (LAPW) and the linearized muffin-tin-orbital (LMTO) methods, usually with some version of the local density approximation (LDA), and without attempting to account for strong correlations. Therefore, the bands were doubly occupied with up- and down-spins to give a half-filled conduction band in which the Fermi level is located. The parent materials were predicted to be metallic. When neutron scattering results [6] clearly showed them to be antiferromagnetic insulators, the need to include strong onsite correlations became apparent. A few spin polarized calculations appeared in this early period, but they were unable to predict the antiferromagnetic ordering. A comprehensive review of the literature on electronic band and cluster calculations up to 1989 has been given by Pickett [25].

The results of conventional band calculations did clearly emphasize the importance of the $CuO_2$ planes and the quasi-2D nature of the electronic structure. They also provided data to which parameterized calculations could be fit to extract values of the various matrix elements of the Hamiltonian.

Development of suitable first-principles methods for treating the electronic structure of highly correlated systems is an active and important research area in its own right. But progress in it has been slow and the demands on computational resources are daunting. One method that has been developed is the so-called LDA+U, which may or may not be considered a first principles method depending on one's viewpoint. In this method, a $U$ term to account for strong onsite correlations is simply incorporated into a local spin density calculation from the beginning, much as in parameterized band and cluster calculations. This approach has been applied to a variety of transition metal materials with good success. A review of this work has been given by Anisimov *et al.* [65]. More recently, LDA has been combined with dynamical mean field theory (DMFT) to treat highly correlated systems [66,67]. Since the results of first principles calculations are not essential to our objectives, they will not be considered further.

Several somewhat similar approaches can be taken to circumvent the difficulties with strongly correlated first-principles calculations on a phenomenological level. In Ref. [42], which we will follow here, a type of calculation introduced by Slater [68] many years ago to study antiferromagnetism in Cr and other $3d$ transition metals was employed. As Herring [15] and others have observed, Slater's approach does not lead to a fundamental theory of antiferromagnetism because it does not, for example, provide a satisfactory treatment of the phase transition at the Neel temperature. Nevertheless, it is useful in many respects and particularly in the way in which correlation is introduced. It also has the advantage of starting from the magnetic configuration known experimentally to exist in the undoped cuprates.

In Slater-type calculations, the chemical unit cell (CUC) is doubled to form a magnetic unit cell (MUC) which contains one magnetic ion with predominantly up spin and one with predominantly down spin; correspondingly, the chemical Brillouin zone (CBZ) is halved to form the magnetic zone (MBZ). In the $3d$ metals considered by Slater, there was a direct interaction between the up- and down-spin sites via an effective exchange interaction. In the case of transition metal oxides, the magnetic ions interact only indirectly through the oxygen ions via superexchange and the occurrence of up and down spins on the same Cu site is inhibited by the introduction of a Hubbard $U$ term. The effect is to open a gap in the density of states (DOS) obtained from a conventional band calculation so that a material such as stoichiometric $La_2CuO_4$ is found to be an insulator (or semiconductor), as observed. Energetically, the up- and down-spin bands are degenerate, as in conventional band theory, but the wave functions for the two spin bands with the same energy are different. There is virtually no "double occupancy" of the Cu sites although the O orbitals retain the double occupancy characteristic of $O^{2-}$.

The mean-field Hamiltonian used for the band calculations can be obtained in a straightforward way from Eq. (3). In fact, Hamiltonians for the $\alpha$- and $\beta$-spin systems are totally independent except is so far as they interact via the $U$ terms. With $s=\alpha$ or $\beta$ and $s \neq s'$, we can write

$$H_s = \sum_{ij,s} \varepsilon_{ij} c_{is}^\dagger c_{js} + \sum_i U_{is,is'} n_{is'} n_{is} \quad . \tag{11}$$

Note again that *i* and *j* run over Cu and O sublattices. In the band calculations described next, three diagonal terms giving the energies of the Cu *3d* ($x^2$-$y^2$) and the O *2pσ* and *2pπ* orbitals were retained from the first sum; we denote these by *ε(d,d)*, *ε(σ,σ)*, and *ε(π,π)* respectively. Four off-diagonal terms, denoted by *ε(d,σ)*, *ε(σ,σ′)*, *ε(π,π′)*, and *ε(σ,π′)* were kept. The first of these is just the hopping integral between the *d* orbital and the nearest neighbor O *2pσ* orbitals. It is frequently the only such integral retained in simplified parameterized band calculations, but there is a substantial body of results for transition metal and alkaline earth oxides that shows the nearest neighbor O-O interactions are generally too large to be neglected. They are of particular importance to the present problem because they may provide hole conduction paths that avoid the Cu sites. Their inclusion is essential in both the Hubbard and t-J models to explain several experimental results, particularly those of the angle resolved photoemission spectroscopy (ARPES) experiments that will be discussed below. Finally, there are the Hubbard terms that give the repulsive energy when *α*- and *β*-spin electrons are located on the same site. We will retain a *U* term only on the Cu sites and neglect those on all O sites.

The "three-band effective Hubbard model" [22] that has been so widely used in the literature focuses on the role of the Cu *d($x^2$-$y^2$)* orbital and the two O *2pσ* orbitals of the three-atom CUC. This is thought to be enough to describe the essential physics related to the superconductivity itself. However, inclusion of the other orbitals is necessary to fully describe the optical, magnetic, photoemission, etc., properties [33].

The following set of parameters (in *eV*) was extracted from conventional band calculations: *ε(d,d)=ε(σ,σ)= -3.00, ε(π,π)= -2.40, ε(d,σ′)= -1.10, ε(σ,π′)=0.40, ε(σ,σ′)= 0.55, ε(π,π′)=0.45*. After some exploratory tests, a value of *U=3.0* was chosen to use with this set of parameters. It gives a band gap of ~*1.5 eV*, close to the observed value. However, the results of the calculations are not particularly sensitive to this choice. In Ref. [42] several other sets of parameters were used and the results compared to establish the sensitivity of the dispersion curves and wavefunctions to some of the parameters.

In fact, a wide variety of values for these parameters has appeared in the literature. This then raises the question of what interpretation the numerical values of the parameters in such a calculation may have, as discussed at length by Pickett in his review article, Ref. [25]. While we cannot address this issue in depth here, the following considerations illustrate the complexity of the problem.

The basis functions for parameterized calculations are often referred to as "atomic orbitals" and those on different sites are usually assumed to be orthogonal. It is then only the matrix elements of the Hamiltonian that are determined by fitting to first principles calculations. The so-called overlap matrix is a unit matrix. Such a procedure was originally introduced by Slater and Koster [69] as an interpolation scheme for the tight-binding method. The atomic orbitals may be considered as approximations to the symmetrically orthogonalized (Löwdin) orbitals or to the Wannier functions which, in principle, could be derived from the band functions if they were known. Although the Wannier and Löwdin functions on different sites are orthogonal, they have oscillations that may extend well beyond nearest neighbors and this generates hopping terms between the corresponding sites. Also, in the solid it is not actually the atomic orbital energies that enter because the effects of Madelung potentials, nonorthogonality, screening etc., must be included. It may be concluded that considerable care should be exercised in attributing a precise physical interpretation to a particular set of parameters.

Dispersion curves in the energy range near the Mott-Hubbard (M-H) gap are shown in the left panel of Fig. 3. The band maxima occur at the *M/2* points with secondary maxima, ~*0.5 eV* lower, along the *Γ-X* direction. At the *M/2* maxima the square of the wave function is *0.29 3d($x^2 - y^2$)*, *0.57* O *$2p\sigma$*, and *0.14* O *$2p\pi$* and at the *X* point it is an equal mixture of *3d($x^2-y^2$)* and O *$2p\sigma$*. The maxima at the *M/2* point is a direct result of the assumed AF order and the Hubbard *U* term.

With the Fermi level for the AF insulator taken at the top of the valence band for convenience, the Fermi surface can be described either as enclosing a large electron pocket centered at the *Γ* point or an equally large hole pocket at the *M* point. As the Fermi surface moves into the valence band with hole doping, pockets open around the *M/2* points in the MBZ (right panel of Fig. 3). Or, alternatively, in the CBZ the hole pocket around *M=($\pi,\pi$)* expands beyond the half-filling volume.

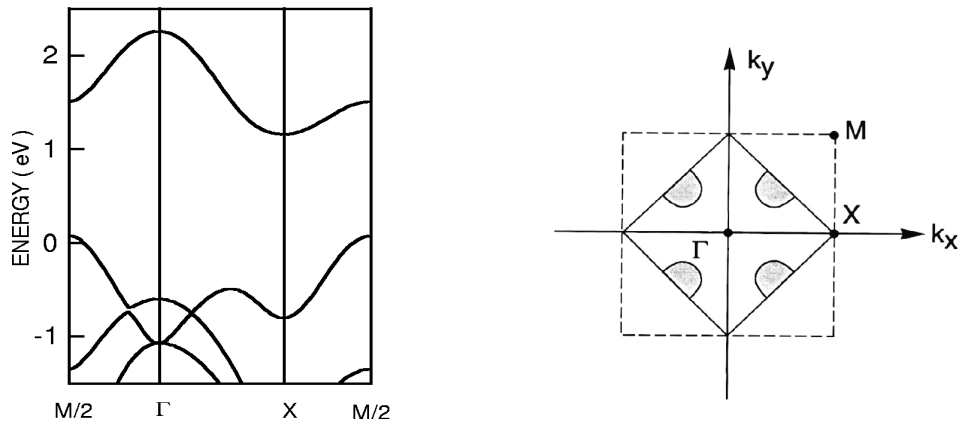

Figure 3. The left panel shows the energy bands in the vicinity of the Mott-Hubbard gap. The right panel shows the Brillioun zone in the MUC with hole pockets due to doping at the *M/2 =($\pi/2,\pi/2$)* point.

Projected density of states curves (in arbitrary units) are given on Fig. 4 by the full lines. The dashed curves show the full DOS, but without the contributions from the O *2pz* and Cu *d(zx)* and *d(zy)*. When these are included and the resulting curve convoluted with a resolution function, quite satisfactory agreement with the experimental photoemission results are obtained. The *d-αα* panel shows the projected DOS for *α*-spins on the *α*-sublattice, while *d-αβ* shows their density on the *β*-sublattice. The latter is very small below the Hubbard gap and large above it, as expected. Both O-$2p\sigma$ and $d$-$\alpha\alpha$ densities are high at the valence band edge.

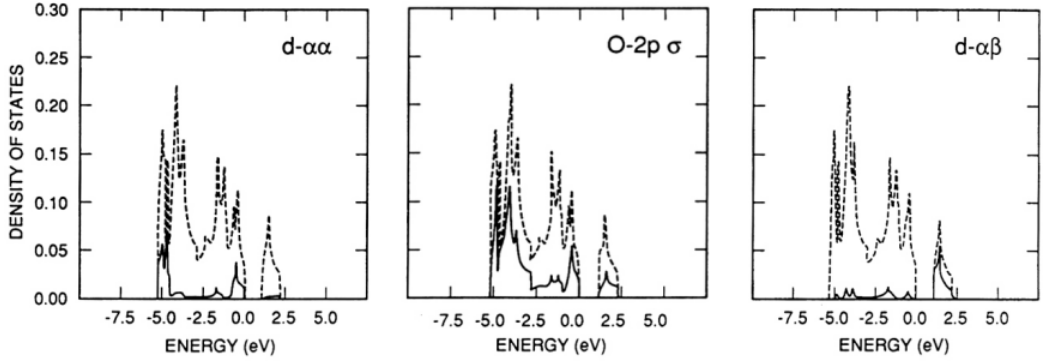

Figure 4. Projected densites of states as described in the text. From Ref. [42].

The calculations show, again as expected, that the M-H gap closes when $U=0$. Since the $\alpha$- and $\beta$-spin bands are completely degenerate, the Fermi level is located by filling each set of bands with one-half the total number of electrons. The Fermi level then falls at the bottom of the M-H gap (top of the valence band), making $La_2CuO_4$ an insulator, as observed; with $U=0$ it would be a metal. Also, with $U=0$ there is no magnetic moment on the Cu ions, whereas moments close to the value of $(0.48 \pm 0.15)\mu_B$ reported in the literature [70-72] are found for $U\sim 3$-$6$ $eV$.

### b.) Holes in the $CuO_2$ planes and spin polaron formation

We assume that for low doping, holes are introduced into the valence band in a manner analogous to the case of nonmagnetic semiconductors and insulators. These holes, however, are dressed by interactions with the Cu spins to form the spin polarons (we do not consider ionic polarons). In a first approximation and one that appears to be surprisingly good, the principal effect of the spin-polaron formation is to increase the effective mass of the carriers whose dispersion curves are given in Fig. 3. In Ref. [42], the spin-polaron effective mass, $m_p$, was assumed to be related to the band effective mass, $m_b$, by an equation from Mott and Davis [46], i.e.,

$$m_p = m_b \exp(\gamma R_p), \qquad (12)$$

as already mentioned in Sec. 2. The polaron radius, $R_p$, is given in units of $a^*$, the nearest-neighbor distance between Cu spins on the same sublattice, and $\gamma$ is approximately unity. This expression was obtained by considering the change in the orientation of the spins at the periphery of the spin polaron, relative to the AF background, as the hole moves through one AF lattice spacing. Clearly, the expression is meant to hold when the radius of the polaron becomes large. However, at $R_p\sim 1$, the effective mass increase due to spin polaron formation is a factor of ~3.0 greater than the band effective mass. This value of $m_p$ is quite close to that needed to fit the dispersion curves for $Sr_2CuO_2Cl_2$, as discussed below. This may be fortuitous, but it does indicate the importance of spin excitations in determining the dispersion curves and bandwidth.

The density of states from the $d$-$\alpha\alpha$ and O $2p\sigma$ panels of Fig. 4 shows a peak ~*1 eV* wide at the top of the lower Hubbard band and this can be traced primarily to the antibonding bands of the two orbitals. Renormalizing the width of this band by a factor of *3* reduces it to *~0.3 eV*, which is almost exactly that found from the ARPES results discussed later.

To go beyond our first approximation, it is necessary to treat holes in an AF insulator in a more rigorous manner. This highly complex problem has been studied by many authors in both the Hubbard and t-J models, especially since the discovery of HTSC. To consider it in any detail would be outside the scope of this article (see, however, Sec. 4 on the t-J model), but two aspects of the problem should be noted.

As mentioned in Sec. 2, Zhang and Rice [26] sought to demonstrate that the three-band Hubbard model could be reduced to an effective one-band t-J model. They emphasized the importance of the spin singlet that can be formed between the spin of an O $2p\sigma$ hole and that of the hole in the $d$-orbital on the neighboring Cu site. It was argued that the binding energy of this singlet is so great that it is able to move through the lattice as an entity. It is effectively a vacancy in the antiferromagnetic background, just as in the case of the one-band t-J model discussed in Sec. 2. This result, although still somewhat controversial, is important because it opens up the extensive theoretical formalism of the simple t-J model to the more complex HTSC systems. It is not difficult to see that in an AF background, if the Zhang-Rice singlet and the spin orientation of the hole are to be maintained as the hole propagates, the Cu spins on one sublattice must be flipped. This is intimately related to the formation of string spin polarons, as we have seen in Sec. 2(c).

Perhaps the single most important conclusion from a large number of studies on both the Hubbard and the t-J models is that the effective bandwidth is of the order of *2J*, or about *0.25 eV*. We have already seen indications of this from the Mott-Davis expression for the effective mass, but it is a quite general result. As might be expected, the width comes not from the localized ferron or string polarons themselves, but from the spin-flip $S^+S^-$ terms in the Hamiltonian which allow for propagation. This will be discussed further in the next section.

The ARPES data of Wells *et al.* [73] for $Sr_2CuO_2Cl_2$ is shown in Fig. 5; the fits to the data given by the curves will be considered shortly. $Sr_2CuO_2Cl_2$ has $CuO_2$ planes quite similar to those in the high-$T_c$ cuprates, but it is difficult to dope so that carrier-free samples are readily obtained. This is crucial for the ARPES measurements on the AF insulating parent material. Consequently, these measurements have become somewhat of a testing ground for studying the dispersion curves of a single hole in an AF spin-*1/2* background. It was pointed out in Ref. [73] that the single-band t-J model could not fit the ARPES data. Specifically, the calculated dispersion about the *($\pi/2,\pi/2$)* point along the

MBZ boundary is far too small and the nearly flat band in the *(0,0)–(π,0)* direction that is observed experimentally is not obtained.

Several groups have addressed this problem. Nazarenko *et al.* [74] studied it shortly after the data of Ref. [73] became available. They demonstrated that the simple t-J fit to the data could be greatly improved by adding an O-O hopping term. Starykh *et al.* [75] considered the problem within the context of an extended Hubbard model using the self-consistent Born approximation (SCBA). They also found that the nearly isotropic dispersion about the *(π/2,π/2)* point came from the inclusion of the O-O hopping integral. A value of *0.76 eV*, together with other parameters in the model gave a satisfactory fit. This is consistent with the value of *0.55 eV* used in the band calculations described in the preceding subsection.

Belinicher *et al.* [76] studied the same problem using an extended t-J model. They introduced properties of orthogonalized orbitals mentioned above to justify O-O hopping terms at distances greater than first nearest neighbors. While this may be a more rigorous way of approaching the problem, it makes it difficult to compare specific parameter values in the various calculations. Nevertheless, the importance of O-O hopping was clearly demonstrated. An illustration of their fit is shown in Fig. 7 of Sec. 4.

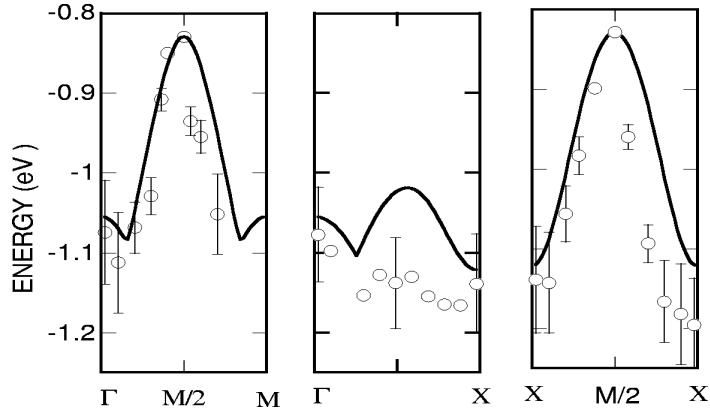

Figure 5. Comparison of the dispersion curves of Fig. 3 with ARPES results for $Sr_2CuO_2Cl_2$.

For the calculations leading to Fig. 5, the polaron effective mass was assumed to be a factor of three greater than the band effective mass. This factor of *~3* is quite close to that extracted from the Mott-Davis expression for the polaron effective mass and used in Ref. [42] for calculating the transition temperature. The bare band dispersion curves from Fig. 3 were scaled accordingly. The fit is remarkable, especially considering that the measurements were made more than five years after the calculations. The agreement could be improved by varying the parameters somewhat. By putting the O-O hopping in the calculations of Ref. [42] equal to zero it is readily established that the calculated dispersion about *(π/2,π/2)* is then highly anisotropic, as the later calculations found.

The close agreement between the results of these four totally independent calculations [42,74-76] and with experiment argues for the essential correctness of both the band structure shown here and the extended Hubbard and t-J models, even though some of the detailed parameters may not seem to agree closely. (This is probably another illustration of the need for caution when interpreting these parameters.)

### c.) Gap Function and Critical Temperature

In this subsection we sketch the derivation of the gap symmetry based on the foregoing model, with consideration that the pairing may go by either (or both) the first and second neighboring cells in the AF lattice. It is hoped that this may clarify the origins of the various gap symmetries that have played such an important role in discussions of high-$T_c$ properties. The transition temperature is also calculated within this framework [42].

The reduced pair Hamiltonian [77] can be written as

$$H_{red} = 2\sum_{\mathbf{k}} \varepsilon_{\mathbf{k}} b_{\mathbf{k}}^{\dagger} b_{\mathbf{k}} + \sum_{\mathbf{k},\mathbf{k'}} V(\mathbf{k},\mathbf{k'}) b_{\mathbf{k}}^{\dagger} b_{\mathbf{k'}} \quad , \tag{13}$$

where $V(\mathbf{k},\mathbf{k'})$ is the pairing matrix element and the operators

$$b_{\mathbf{k}}^{\dagger} \equiv c_{\mathbf{k}\uparrow}^{\dagger} c_{-\mathbf{k}\downarrow}^{\dagger} \,, \qquad b_{\mathbf{k}} \equiv c_{-\mathbf{k}\downarrow} c_{\mathbf{k}\uparrow} \tag{14}$$

create and destroy, respectively, singlet Cooper pairs. As discussed above, the $\varepsilon_k$ are considered to be given to a first approximation by the hole energies near the top of the valence band, renormalized by the induced spin deviations, i.e., they are the single spin-polaron energies in the Bloch representation. In order to find the gap function, it is necessary to simplify $H_{red}$. This is commonly done by making a Hartree-Fock-like approximation to reduce the product of four single-particle creation and annihilation operators to a product of two. In this way, the gap parameter

$$\Delta(\mathbf{k}) = \sum_{k'} V(\mathbf{k},\mathbf{k'}) \langle b_{\mathbf{k'}}^{\dagger} \rangle_s = \sum_{k'} V(\mathbf{k},\mathbf{k'}) \langle c_{\mathbf{k'}}^{\dagger} c_{-\mathbf{k'}\downarrow}^{\dagger} \rangle_s \tag{15}$$

is introduced, where $\langle \,\rangle_s$ indicates an average over pair states in the superconducting ground state. To calculate this average, the $b$'s must be determined by diagonalizing $H_{red}$, so that a self-consistency requirement is introduced.

From general requirements [78], for singlet pairing the gap function must have even parity in $\mathbf{k}$, i.e., $\Delta(-\mathbf{k}) = \Delta(\mathbf{k})$. We write

$$V(\mathbf{k},\mathbf{k'}) = \int d\mathbf{r}_1 d\mathbf{r}_2 \, V(\mathbf{r}_1 - \mathbf{r}_2) \psi_{\mathbf{k},\alpha}^*(\mathbf{r}_1) \psi_{-\mathbf{k},\beta}^*(\mathbf{r}_2) \psi_{\mathbf{k'},\alpha}(\mathbf{r}_1) \psi_{-\mathbf{k'},\beta}(\mathbf{r}_2) \quad , \tag{16}$$

and take

$$\psi_k(r) = (1/\sqrt{N}) \sum_\mu \exp(i\mathbf{k}\cdot\mathbf{R}_\mu) \phi_k(\mathbf{r}-\mathbf{R}_\mu) \ . \quad (17)$$

$R_\mu$ is a lattice vector in the magnetic lattice and $\phi_k(\mathbf{r}-\mathbf{R}_\mu)$ is one of the basis functions. The resulting expression can be simplified using translational invariance and by making the assumption that the overlap charge density of the orbitals on different sites is small compared to the site-diagonal value, i.e.,

$$\phi^*_k(\mathbf{r}-\mathbf{R}_\mu)\phi_k(\mathbf{r}-\mathbf{R}_\nu) \sim 0 \ \text{if} \ \mathbf{R}_\mu \neq \mathbf{R}_\nu \ . \quad (18)$$

Also, results in Ref. [42] suggested that in 2D the singlet pairing potential is sharply peaked when the two holes are at distances $a^*$ and $\sqrt{2}a^*$ from each other. Here we will simplify the calculation by assuming that $\mathbf{R}_\mu$ can run only over ions in the first and second neighbor magnetic unit cells from the origin. Also, $\mathbf{R}_\mu$ is assumed to be a unit cell vector, thus neglecting the difference in phase factors between the different sites within a cell. Then, with $\mathbf{q} \equiv \mathbf{k}-\mathbf{k}'$

$$V_{k,k'} = (V_1/N)\sum_{1nn} \exp(i\mathbf{q}\cdot\mathbf{R}_\mu) + (V_2/N)\sum_{2nn} \exp(i\mathbf{q}\cdot\mathbf{R}_\mu) = V_1(\mathbf{k},\mathbf{k}') + V_2(\mathbf{k},\mathbf{k}') \ , \quad (19)$$

where

$$V_1(\mathbf{k},\mathbf{k}') = (2V_1/N)(\cos q_x a^* + \cos q_y a^*) \ ,$$
$$V_2(\mathbf{k},\mathbf{k}') = (4V_2/N)\cos q_x a^* \times \cos q_y a^* \ . \quad (20)$$

By expanding the cosine terms, keeping only the even parity components, and projecting with functions that transform as the various irreducible representations, four pairing potentials are obtained. It was found that $V_1(\mathbf{k})$ decomposes into the two components

$$V^s_1(\mathbf{k}) = A(\cos k_x a^* + \cos k_y a^*) \ , \qquad V^d_1(\mathbf{k}) = B(\cos k_x a^* - \cos k_y a^*) \ , \quad (21)$$

and $V_2(\mathbf{k})$ into the two

$$V^s_2(\mathbf{k}) = C_{xy}\cos(k_x a^*)\cos(k_y a^*) \ , \qquad V^d_2(\mathbf{k}) = S_{xy}\sin(k_x a^*)\sin(k_y a^*) \quad (22)$$

The gap equation can be similarly decomposed into

$$\Delta(\mathbf{k}) = \sum_n f_n(\mathbf{k})\Delta^0_n \ , \quad (23)$$

and a secular determinant for the $\Delta^0(n)$ written. However, the elements of this equation are non-linear functions of the $\Delta^0(n)$ themselves and cannot be solved without further simplifications [79]. As a consequence of this non-linearity, the general solution of Eqs.

(15) and (23) is given by a mixture of states of different symmetries and cannot be classified as pure *s*-like or pure *d*-like as is so often done. However, one can hope that one symmetry or the other predominates and this is indeed believed to be the case [79,80]. For the high-$T_c$ materials, there seems to be considerable evidence [80] that the gap is predominantly of $d(x^2-y^2)$ symmetry (see Sec. 4).

By making a series of simplifying approximations, calculations of the gap as a function of *x* for the four potentials just described were carried out in Ref. [42]. These approximations took the hole pockets to be circular about the *M/2* points, used an effective mass approximation for the energies and wave-functions, and assumed that sums over the Fermi surface could be broken up into sums over the individual pockets. The most interesting aspects of these calculations is that a good fit to the experimental data for $T_c(x)$ could be obtained intrinsically, based solely on the properties of the Fermi surface in the MBZ. This is shown in Fig. 6.

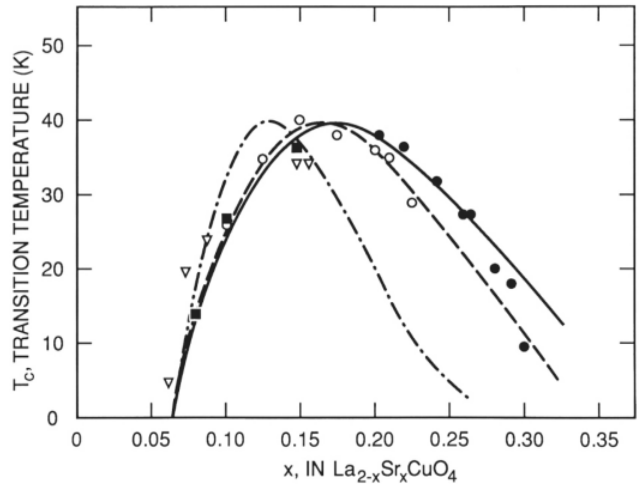

Figure 6. Transition temperature as a function of doping as described in the text. From Ref. [42].

The experimental data on the figure are taken from the following sources: empty circles - Tarascon *et al.* [81]; triangles - Schafer *et al.* [82]; squares - van Dover *et al.* [83]; filled circles - Torrance *et al.* [84]. The three calculated curves on the figure were obtained by making various assumptions about the symmetry and strength of the pairing potential, as discussed in Ref. [42]. For example, the solid curve is the result for the $V_1^d$ (~ $d(x^2-y^2)$) symmetry (holes in nearest neighbor cells of the magnetic lattice).

We do not need to describe the results on Fig. 6 in detail but it can be seen that the $T_c(x)$ curve peaks for *x~0.15* and then declines to *0* at *x=0.30-0.35*. The rise in the underdoped region occurs because more **k**-space becomes accessible for pairing as the Fermi surface moves away from the MBZ boundary. The curve begins to fall after the maximum because of the fall-off of certain form factors in the calculation which decrease rapidly as the Fermi surface moves away from the boundary of the MBZ.

In spite of the apparent excellent agreement with experiment for some of the curves, it must be kept in mind that the long-range AF ordering on which the Fermi surface calculations were made has been suppressed even at modest doping for real systems. See the discussion of Fig. 13 in the next section. The ARPES results for $La_{2-x}Sr_xCuO_4$ [33] may suggest that the schematic Fermi surface in Fig. 3b is not correct for doping levels much greater than a few percent. However, the interpretation of ARPES experiments is still an active area of research and consensus conclusions have yet to be reached for several of the cuprates [85,86]. We now turn to a consideration of the t-J model.

## 4. t-J MODEL: SPIN POLARONS, PAIRING, SUPERCONDUCTIVITY, AND STRIPES

### a.) Single-hole problem

The nature of the charge carrier in doped Mott insulators has attracted considerable attention in the context of high-$T_c$ physics. In the t-J model, which is closely related to the realistic low-energy model for the cuprates and is also the simplest generic description of doped antiferromagnets, the charge carrier is a spin polaron, as we briefly discussed in Secs. 2 and 3. According to the spin-polaron idea, the hole in its movement disturbs the magnetic background. This can be formally described as the strong coupling of the hole and spin degrees of freedom. This makes the problem similar to the well-known strong coupling electron-phonon polaron problem. However, in spite of the qualitative similarity of these two polarons, there is an essential difference between them. If the phonon polaron can be considered as an almost static object composed of the shifted ions with the electron in the center, the spin polaron is a "spin-bag" with a moving hole inside [52]. Theoretical studies of the t-J model have resulted in a clear understanding of the nature of the low-energy excitations for the system near half filling. The AF spin-polaron concept put forward by Bulaevskii *et al*. [18] was developed in a number of more recent papers [40,47,48,87-97] using different techniques. The main conclusion of these works was that the spin polaron in an AF background is a well-defined quasiparticle with a nonzero quasiparticle residue and a specific dispersion law. The dressing of the hole leads to a narrow quasiparticle band with a bandwidth $W\sim 2J$ for realistic $t>J$, band minima at $k = \pm(\pm\pi/2,\pi/2)$, and a heavy effective mass along the MBZ boundary. The single-hole problem has been treated analytically in detail [47,88-97]. Grouping these efforts, two approaches in treating this problem were used: *(i)* the self-consistent Born approximation (e.g., see Refs. [88-92,95]), and *(ii)* the so-called "string" approach (e.g., see Refs. [47,48,96,97]). A relationship between these two has been established [98] recently. The SCBA method utilizes a property of the hole-magnon interaction, namely the absence of the lowest order correction in the Born approximation series for the single-particle Green's function [88-92,95]. An attractive feature of the SCBA approach is that a single-hole spectral function can be evaluated quite easily using simple numerical calculations. The detailed structure of the single-hole ground state and different current correlations have been studied using SCBA [99,100].

The single-particle properties in the t-J model have been extensively studied by numerical methods as well [8,53]. Exact diagonalization (ED) numerical studies of the t-J model are performed on small clusters with periodic boundary conditions. They are an important source of unbiased information on the low-energy physics of this system. One- and two-hole ground states have been investigated in great detail on the *16-, 18-, 20-, 26-,* and *32*-site clusters [101-126]. For an earlier review of these numerical works, we refer to [8]. More recent numerical studies involved the density-matrix renormalization group approach for much larger clusters, with open or mixed boundary conditions [49], and Monte Carlo calculations in clusters as big as 32×32 sites at very low temperature [127]. These numerical results on the single-hole problem show that the quasiparticle peak at the bottom of the spectral function survives in the thermodynamic limit [127]. The corresponding quasiparticle band is narrow (of the order of *2J* in the "physical" region *t>J*) and the band minima are at the MBZ boundary. These numerical studies indicated that the effective mass around the hole minima is anisotropic in the simple t-J model, and that the quasiparticle residue is reduced substantially for realistic *t>J*, all in excellent agreement with analytical predictions. Further, the full dispersion relation predicted by analytical work based on the SCBA is found to be in excellent agreement with ED on the 32-site cluster [128]. In Refs. [55,129-132], it was shown that the other analytical results of the t-J model, such as correlation functions and finite-size scaling ansatzes, quantitatively reproduce all essential features of the ED cluster data. This gives a strong support for the adequacy of the spin-polaron picture in the t-J model.

Since the analytical and numerical results for the single-hole problem have been reviewed many times [8,133] we would like to restrict ourselves here to the discussion of a few qualitative aspects of the problem and also to focus on the features which, in our view, are important for the further development of the theory.

The angle-resolved photoemission experiments on insulating $Sr_2CuO_2Cl_2$ [73], as already discussed in Sec. 3, can be considered a direct test for a single-hole dispersion relation within the low-energy models of the $CuO_2$ plane. They have been described successfully within the t-J-like model, as shown by several studies [54,74-76,134-136] (see Sec. 3 for more details). The experimentally observed dispersion relation, $E_k$, for a single hole has the following characteristic features [73]: *(i)* bandwidth about $W\sim 2J$, *(ii)* band minimum at the $k=\pm(\pm\pi/2,\pi/2)$ points, *(iii)* isotropic dispersion near the band minimum, and *(iv)* almost flat dispersion along the line $(0,0)\rightarrow(\pi/2,0)\rightarrow(\pi,0)$. The first two results agree with the t-J model spin-polaron calculations discussed above. For explaining the isotropic dispersion around the band minima and the flat regions on the top of the hole band, extra terms must be included in the t-J model. Such terms, which include the next- (*t'*), next-next-nearest-neighbor (*t''*), and the so-called "three-site" hopping integrals ($t_S$, $t_N$), should be present in the realistic low-energy model of the $CuO_2$ plane as suggested by first-principle calculations [137]. These additional terms provide an adequate description of the experimental data within the framework of the t-t'-t''-J model. Fig. 7 shows an excellent agreement with experiment along the main BZ directions.

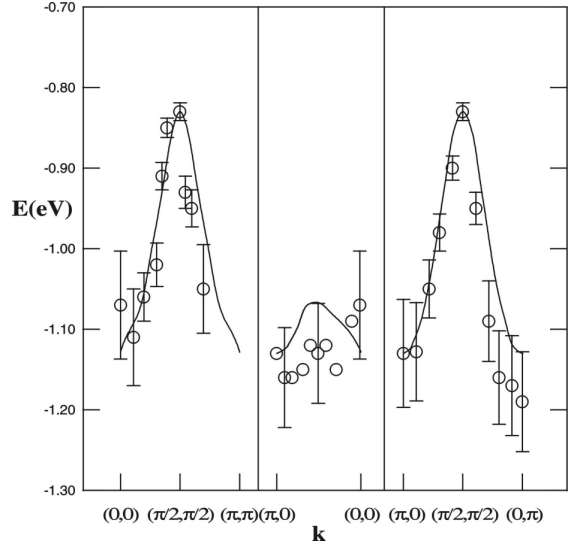

Figure 7. Dispersion curve of a hole in the generalized t-t'-J model [76], along the main BZ directions $(0,0) \to (\pi,\pi)$, $(\pi,0) \to (0,0)$, and $(\pi,0) \to (0,\pi)$ (solid curve). Model parameters that provide this $E_k$ are: $t/J=2.5$, $t'=-0.2t$, $t''=0.15t$, $t_S=-2t_N=-J/4$, $J=0.14$ eV (see Ref. [76]). Experimental results from Ref. [73] are also shown (open circles), (after Ref. [76]). Note that the electronic energy is shown, therefore the top of the curves correspond to the bottom of the hole band.

Two remarks are necessary here. It can be shown that, in fact, the quasiparticle bandwidth $W \sim 2J$ is a universal prediction of all possible generalizations of the t-J model in the region of parameters when $t>J$. The reason for such universality has been first outlined by Kane *et al.* [87]. Roughly speaking, it is similar to the Cherenkov effect: due to the energy and momentum conservation there is a threshold energy below which a massive quasiparticle cannot emit an acoustic excitation with linear dispersion. Therefore, such a quasiparticle must be well defined below such energy. In addition, if the coupling between the massive and acoustic quasiparticles is strong, there are no coherent excitations *above* that threshold energy. In the t-J-like models, the threshold energy is set by the magnon dispersion $\sim 2J$, and the coupling is set by the nearest-neighbor hopping $t$. This means that the coupling is always strong if $t>J$. Thus, one should expect to see a strongly renormalized narrow coherent band with the width defined by the magnon bandwidth ($\sim 2J$) at the bottom of a much wider incoherent band ($\sim 8t$, width of the "bare" band), see Fig. 8 for an illustration. From these physical arguments, it is clear that in the presence of strong coupling to the spin fluctuations there are no coherent quasiparticles at energies higher than $2J$. The basic arguments are the absence of the hole-magnon scattering near the bottom of the band and its domination at higher energies.

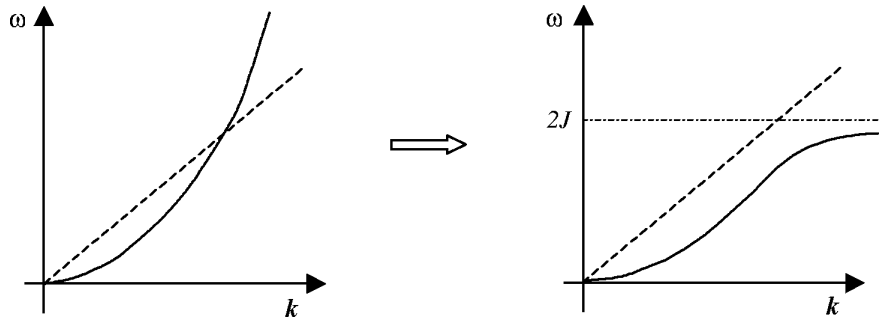

Figure 8. A schematic picture of the massive (solid line) and massless (dashed line) modes in the case of no coupling (left) and in the case of strong coupling (right). Dashed-dotted line shows the threshold energy.

Notice that another feature of the ARPES results can be explained within the same qualitative picture. At the top of the hole band, the intensity of the quasiparticle peaks should be suppressed since that energy is close to the threshold energy and much of the weight should be transferred to the incoherent part of the spectrum. Analytical calculations within the t-t'-t''-J model for the states at the top of the band show a small quasiparticle peak and a rather wide incoherent subband right next to it [76,95]. It immediately explains the width of the broad peaks in the ARPES energy-distribution curves for the *k*-points at the top of the hole band, shown in Fig. 7 by the error bars.

The second remark concerns the sensitivity of the calculated hole "band structure" to the parameters of the t-t'-t''-J model. In fact, a concern was raised that the necessity to "fine tune" the parameters of the model in order to fit the experimental data signifies an ideological failure of the spin-polaron concept [138]. As we discussed above, the width of the quasiparticle band is set by the magnon bandwidth and is a universal prediction of the model. The position of the minima at $k = \pm(\pm\pi/2, \pi/2)$ points does not restrict any parameter, but only requires the *t'* term to be negative. This latter appears to be a natural consequence of the kinetic energy minimization requirement. The concern is the *isotropy* of the dispersion around the band minima, which *is not* a property of the "bare" t-J model. However, an important feature of the energy spectrum of the generalized t-t'-J model has been found in Ref. [76]: if the values of *t'* terms are not too small ($|t'| + |t''| + ... > J/2$), the *shape* of the quasiparticle band is almost insensitive to the changes in these parameters. The variations of these parameters in the limits which definitely cover the parameters ranges provided by the first-principle calculations, affect only the reference energy and the quasiparticle residues. This feature is easy to understand: when *t'* terms are not too small, they already form the "bare" band, $W_0$, which is wider than the characteristic energy *2J*. Then, the further increase of *t'* concerns only the states higher than *2J* in that band, which become incoherent due to magnon emission according to the above discussion. Therefore, one can expect a rather universal shape of the single-particle band in all $CuO_2$ materials.

Since much of our knowledge about the properties of the t-J model is gathered from numerical studies, one would wish for an analytical theory which agrees with the numerical data on all essential points, thus providing a definite physical answer on how the ground state of the system is formed and what are the excitations around that state. The self-consistent Born approximation as well as some other methods applied to the t-J model within the spin-polaron approach have provided a quantitative agreement with the ED numerical data on the single- and two-hole properties. However, application of this method to the many-hole problem is technically more challenging and cumbersome, so that how to ascribe the results to a definite physical behavior is not always clear. For a comparison of the numerical and analytical results, the substantial finite-size effects in the small clusters can be an issue too. For the sake of further discussion, we consider here in more detail a limiting case of the t-J model, the so-called t-$J_z$ model. For this model, an analytical solution of the single-hole problem is known [56], and it provides very close agreement with the numerical data. Also, the finite-size effects in the clusters are much weaker for the t-$J_z$ model. More importantly, a detailed analysis of the pairing mechanism and of stripe formation can be made within the framework of this model. The Hamiltonian of the t-$J_z$ model is:

$$H = -t\sum_{\langle ij\rangle\sigma}\left(c^\dagger_{i\sigma}c_{j\sigma} + H.c.\right) + J\sum_{\langle ij\rangle}\left(S^z_i S^z_j - \frac{1}{4}n_i n_j\right) \ , \qquad (24)$$

which differs from the t-J model, Eq. (2), by the absence of the spin-flip terms in the exchange energy. One may ask if the conclusions obtained from the studies of the anisotropic t-$J_z$ model are valid for the isotropic, *SU(2)* symmetric t-J model of Eq. (2). The answer is based on the following observation: the short-range spin excitations in both models are very similar since they correspond to the spin flips, while the long-range excitations in the t-J model are gapless Goldstone modes. Therefore, as long as one is concerned with the short-range physics, these models should lead to similar results. This can be illustrated by the following example. As we discussed in Sec. 2, the single-hole ground-state energy has a characteristic *2/3*-power law dependence on the exchange energy *J* (see Eq. (10)): $E_{GS}/t \cong -2\sqrt{3} + A\cdot(t/J)^{2/3}$ which is observed in the numerical data from finite clusters for both the t-J and the t-$J_z$ model [8]. This argues that the "string" picture for the spin polarons should work equally well in both models. Physically, this means that when the hole motion is fast and the spin relaxation is slow ($t \gg J$) the hole moves in the background of essentially static, staggered spins, as argued in Sec. 2.

We provide here a *quantitative* comparison of the ground-state energies for the single-hole states in the t-J and t-$J_z$ models obtained from the 32-site cluster, Fig. 9 [139]. It shows that both the functional dependence and the absolute values of the ground-state energies are very close in the two models. The stronger finite-size effects in the t-J model are the reason for the larger discrepancy at larger *t/J* values.

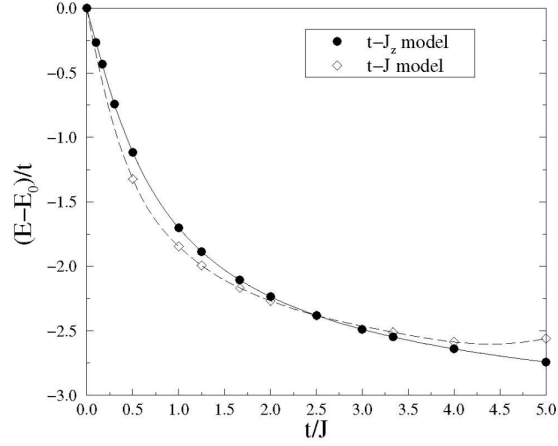

Figure 9. Ground-state energy of the single hole in a *32*-site cluster in the t-$J_z$ (circles) and t-J (diamonds) models as a function of *t/J*. $E_0$ is the ground-state energy of the *32*-site cluster with a static hole in it (*t=0*). Lines are guides to the eye.

In Sec. 2(c), we briefly outlined the "string" approach to the spin-polaron problem within the t-$J_z$ model. It includes two steps. First, the problem of the hole motion in a square lattice AF is mapped onto the discrete Schrödinger equation for the motion in the Bethe lattice, with the first branching *z=4* and all other branchings *z-1=3*. At this step the so-called Trugman paths [97] and effects of self-tangencies of the hole trajectory on the energy of the string are neglected. Second, the continuum approximation is applied, which makes the problem explicitly equivalent to the problem of a particle in a 1D linear potential, Eq. (9). This immediately provides $J^{2/3}$-dependence of the ground-state energy. The concern here is that the consistency of the continuum approximation requires the average length of the string to be large $\langle L \rangle \propto (t/J)^{1/3} \gg 1$. This invalidates the applicability of the continuum approximation in the physical *t/J~3* range. However, an analytical solution of the original *discrete* Schrödinger equation for the hole motion in the Bethe lattice valid for *any t/J* was found recently [98,56]. According to this solution, the single-particle Green's function for the t-$J_z$ model is given by:

$$G(\omega) = \frac{1}{\omega - 4t^2 G_a(\omega - 3J/2)} \quad , \qquad (25)$$

where the auxiliary function $G_a$ is given by a continued-fraction form:

$$G_a(\omega) = \frac{1}{\omega - 3t^2 G_a(\omega - J)} \quad , \qquad (26)$$

which can be solved using an ansatz:

$$G_a(\omega) = -\frac{1}{\sqrt{3}t}\frac{Y(\omega)}{Y(\omega+J)} \quad . \tag{27}$$

This transforms Eq. (26) into a difference equation

$$Y(\omega+J) + Y(\omega-J) = -\frac{\omega}{\sqrt{3}t}Y(\omega) \quad , \tag{28}$$

which is the recursion relation for the Bessel functions. Thus, the auxiliary function $G_a$ is given by:

$$G_a(\omega) = -\frac{1}{\sqrt{3}t}\frac{J_{-\omega/J}(2\sqrt{3}t/J)}{J_{-\omega/J-1}(2\sqrt{3}t/J)} \quad , \tag{29}$$

where $J_v(x)$ is the Bessel function. Equations (25), (26), and (29) define the single-hole Green's function.

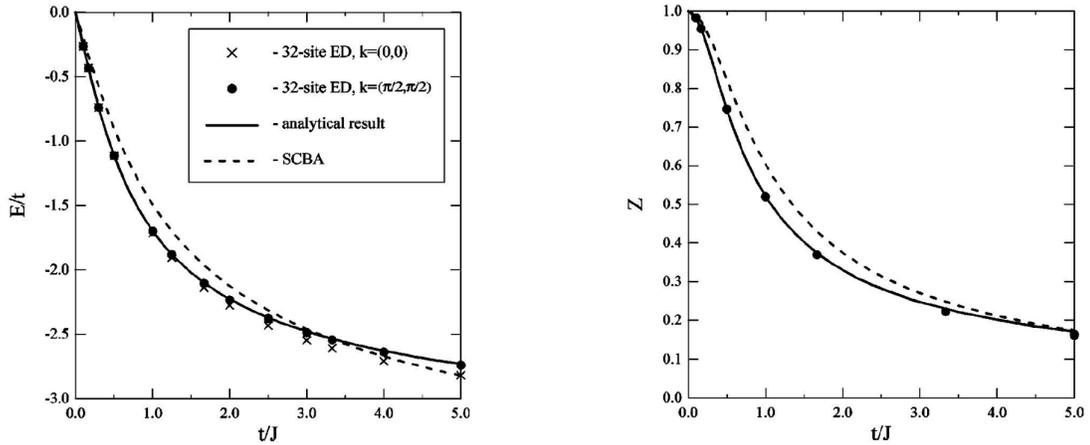

Figure 10. The single-hole ground-state energy (left) and quasiparticle residue $Z$ (right) vs $t/J$. Dots and crosses are the ED numerical data from the *32*-site cluster. Solid curves are the results of the present approach. Dashed lines are the SCBA results, (after Ref. [56]).

Our Fig. 10 shows a comparison of the ground-state energy and the quasiparticle residue $Z$ for the analytical results obtained from Eqs. (25)-(29) and exact diagonalization numerical results for the t-$J_z$ model in the *32*-site cluster with periodic boundary conditions. One can see quantitatively very close agreement of the two in a wide region of $t/J$. Note that the finite-size effects in ED are very small in this case as estimated from a comparison with the other numerical methods, see [56]. In fact, the discrepancy of the analytical results with the numerical data is of the order of the numerical accuracy of the ED.

Such a close agreement of analytical theory and numerical data shows the deep understanding of the spin-polaronic nature of the hole excitations in the t-J model. Using this knowledge, the problems of pairing and stripe formation can be pursued.

### b.) Two-hole problem, pairing

The nature of the pairing mechanism remains one of the central issues in the problem of high-temperature superconductivity. Unusual *d*-wave symmetry of the order parameter, absence of a significant isotope effect, and the existence of a pseudogap, all argue for the secondary role of the conventional phonon-mediated pairing. The magnetic pairing mechanism due to spin fluctuations has been extensively discussed in many theoretical approaches as a main source of unconventional properties of the HTSC materials [8-10]. The investigation of interactions between quasiparticles in the t-J model is indispensable in this context since it is in the microscopic studies of the t-J and Hubbard models where one expects to find answers as to what gives rise to the *d*-wave symmetry of the pairing and on to the origin of the binding forces.

The individual charge carriers in these models are spin polarons, i.e., strongly dressed quasiparticles. Therefore, the problem is to find an effective interaction between these elementary excitations. The signatures of such interactions were sought in the extensive numerical studies of two holes in the finite clusters [8]. It was soon discovered that the holes tend to form a bound state of the *d*-wave symmetry, proving the ability of the t-J model to explain HTSC. Note here that the existence of a bound state in the two-body problem *is not* a necessary condition for the superconducting instability, since the latter comes about as result of many-body effects. Nevertheless, the true *d*-wave bound states in the ground state of the finite clusters have shown the efficiency of the *d*-wave pairing channel in the t-J and Hubbard models and also suggested that the "preformed pairs" mechanism of superconductivity (akin to bi-polaronic superconductivity) might be at work.

The importance of the two-hole problem as the simplest problem that allows one to study interactions between charge carriers in an AF background has been also appreciated in a number of analytical and numerical studies [140-144]. These theories provided two key interactions leading to pairing in the t-J model. One of them is the effective hole-hole static attraction due to minimization of the number of broken bonds by two holes at nearest-neighbor sites (sometimes referred to as the "sharing common link effect"). The other is due to the spin-wave exchange and leads to a dipolar-type interaction between holes. It is the latter which is responsible for the selection of *d*-wave symmetry, while the former enhances the pairing [56].

These works on the low-energy physics of the two-hole system generally describe it in terms of moderately interacting spin polarons [145-147]. There are two questions one may want to ask: *(i)* how generic the *d*-wave pairing is, and *(ii)* is it a robust property of all reasonable generalizations of the t-J model (like t-t'-...-J model)? One may also ask first a simpler question: why is the *s*-wave pairing channel suppressed? The answer to the latter is well understood. Roughly, when the spin wave is a mediator of interaction instead of the phonon there is always repulsion in the *s*-wave channel and, therefore, higher partial waves become favorable instead. In the t-J model, vertex corrections are suppressed and such a repulsion is strong in the physical range of parameters. This generic reason for the absence of the *s*-wave pairing in the magnon-mediated exchange is schematically shown in Fig. 11.

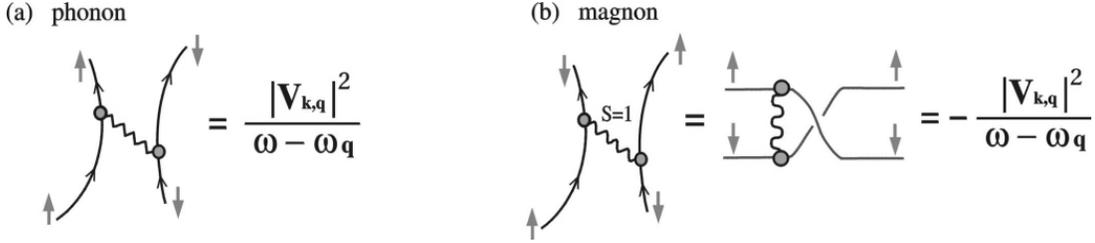

Figure 11. (a) An elementary diagram leading to an attraction due to a phonon exchange. (b) The same diagram for the case of spin-wave exchange. In the latter case spins of electrons are interchanged as a result of exchange and, therefore, the interaction is repulsive. In the t-J model $V_{kq} \sim t$ while $\omega_q \sim J$. This leads to a strong repulsion in the *s*-wave channel when $t \gg J$.

This argument is largely model-independent and is applicable to many other physical systems where magnon-like excitations can mediate interaction between the charge carriers [9,10]. The choice of the higher partial wave for the orbital part of the two-particle wave function allows the repulsion to be canceled out, or even to be turned into an effectively attractive interaction.

However, the question remains why the *d*-wave symmetry is selected over other higher partial waves. The answer to that question is rather subtle. We demonstrate that using an example of the two-hole problem in the t-$J_z$ model. This problem has been thoroughly examined by means of analytical diagrammatic study supplemented by the numerical ED results in the *32*-site cluster [56]. First of all, for the system of two holes in the physical region $t/J > 1$ there are bound states of both *p*- and *d*-wave symmetry. Although the energies of these states are quite close, the ground state is found to be a *p*-wave bound state, not a *d*-wave one, see Fig. 12. From a careful analysis of the interactions involved in the pairing problem it was established that *p*- and *d*-wave bound states would remain degenerate if the hole dispersion were to be neglected. In the t-$J_z$ model, such a dispersion comes from higher order hopping processes [97] and is, in fact, very small. However, it is responsible for the splitting between the *p*- and *d*-wave states shown in Fig. 12. The same figure also shows that the *s*-wave bound state, which is due to the nearest-neighbor "common-link effect", is destroyed at a very small $t/J \sim 0.3$, in agreement with the above discussion. The role of transverse fluctuations in the perturbative limit $J_\perp/J \ll 1$ has been considered in Ref. [56] as well. The fluctuations lead to an anisotropic hole dispersion and to an additional hole-hole interaction. It was shown that both the transverse fluctuations and anisotropic dispersion favor a *d*-wave bound state.

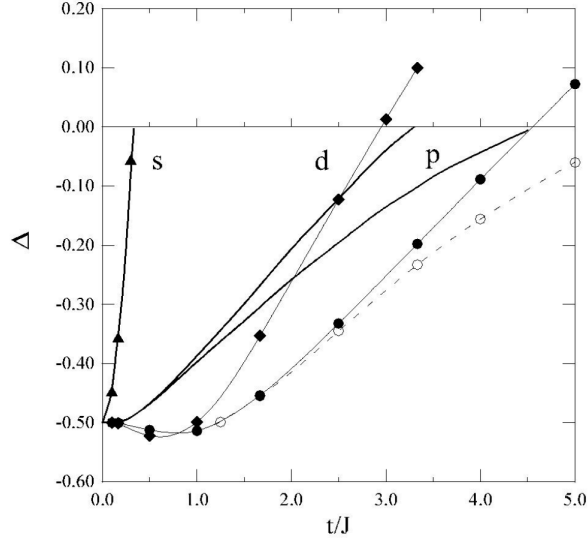

Figure 12. Binding energies of the *s*-, *p*-, and *d*-wave bound states in the t-$J_z$ model *vs* $t/J$. Solid lines are the analytical solutions of the Bethe-Salpeter equation for each partial wave. Triangles, diamonds, and solid circles are the *32*-site ED results for the *s*-, *d*- and *p*-wave, respectively. Lines connecting these points are guides to the eye. Empty circles are the results of a modified Lanczos study on a *50*-site cluster, Ref. [148]. From Ref. [56].

Extensive numerical as well as analytical studies of the realistic case of the isotropic *SU(2)* spins ($J_\perp/J = 1$) in the t-J model have demonstrated a ubiquitous *d*-wave symmetry of the bound state of two holes in the ground state of the system [8,149,150]. We would like to note here that the bound states found in all these studies are rather short-range, a few lattice constants in diameter. Therefore, the short-range interactions, which involve magnons at large ***q***-momenta and at $\omega_q \sim J$, are the most effective for pairing. One may conclude then that the discussed mechanism is effective irrespective of the presence of long-range order. In fact, the ED studies of the t-J model in small clusters do preserve the full *SU(2)* symmetry of the problem and thus do not imply any long-range ordered state. They rather represent a system with well-developed short-range AF correlations. Their presence seems to be the only requirement for the spin-wave exchange mechanism to operate.

As we discussed in the context of experimental ARPES data for the single-hole problem, a realistic low-energy model of $CuO_2$ planes should include *t'*, *t''*, etc. terms. The effect of additional terms on pairing has recently received considerable attention

[151-153]. Since these terms allow the holes to hop within the same sublattice, they should not modify significantly the hole interaction with spins, but should mainly change the single-hole kinetic energy. The earlier analytical and recent numerical studies have shown, however, that the *d*-wave bound state in the two-hole problem disappears if *|t'/t| > 0.1* [142,153]. This is puzzling because in two dimensions arbitrarily small attraction between particles leads to an *s*-wave bound state [154]. Therefore, one may think that if there are no bound states this should indicate that all interactions become repulsive. However, this naive expectation is only valid for the *s*-wave pairing and smooth potentials, and it does not hold for the higher harmonics. In other words, a critical value of the coupling relative to the kinetic energy is required for the *d*-wave bound state to appear even in two dimensions. The *t'*-terms simply shift the balance between the attractive energy and kinetic energy. Therefore, bound states may not exist in the ground state of the two-hole problem in two dimensions, which disqualifies the Bose-condensation of preformed *d*-wave pairs scenario of HTSC, but this does not mean the absence of attractive *d*-wave interactions in the system. Moreover, we expect these interactions to be largely independent of the *t'* terms and to remain active for *d*-wave superconductivity.

As a concluding remark for this subsection, we note that a recent numerical study of the t-J and t-t'-J models in the *32*-site cluster has focused on the problem of the robustness of the *d*-wave bound state [153]. It was found that there are other bound states with finite momenta, namely the *p*-wave states at ($\pi,\pi$) and ($\pi,0$), which are low enough in energy and even become the ground state at *t/J≥3*, close to the physical range for the cuprates. The spatial hole-hole correlations within these states are very different from equivalent correlations within the *d*-wave state. These finite-momentum *p*-wave states could be seriously influenced by the finite-size effects of the cluster and it remains unclear what they would correspond to in the thermodynamic limit. An interesting feature of these states is that they are almost insensitive to the introduction of *t'* terms into the model. This shows that the final word in the problem of pairing in the t-J model is yet to come.

### c.) Superconductivity

Much of theoretical work on superconductivity in high-$T_c$ materials has been done using the t-J model. These efforts included, for example, a variety of mean-field-like approaches which emphasized the importance of different magnetic phases [155]. Here we focus on the quasiparticle-like approaches and on the efforts which explicitly or implicitly involved spin-polarons.

Numerical studies in the finite clusters were used to find indications of superconductivity in the ground state of the t-J model. Note that these studies are different from the problem of the hole binding because the *superconducting correlations* are searched for. In the earlier works by Dagotto and Riera [156] and Dagotto *et al.* [157] the pair-pair correlation function $C(m) = 1/N \sum_i \langle \Delta_i^\dagger \Delta_{i+m} \rangle$ and its susceptibility have been investigated near the quarter-filling density $\langle n_e \rangle \approx 1/2$. Strong signals of *d*-wave superconductivity were observed close to the phase-separation part of the phase diagram.

The role of proximity to the phase separation region was recently discussed in Ref. [158]. The signatures of the superconducting, off-diagonal long-range order were sought in the numerical studies since the presence of such correlations would directly indicate the tendency towards a superconducting state. Such correlations have been found in finite clusters assuming the BCS character of the low-lying states [150]. It was shown that the pairing occurs primarily in the $d(x^2-y^2)$ channel with the corresponding gap value $\Delta_d \cong 0.15 \div 0.27 J$ for the doping between $x=10-50\%$. Recently, a similar study has investigated the problem in larger clusters and with more focus on the $\boldsymbol{k}$-, $\omega$-structure of the $d$-wave gap function [159].

Analytical investigations of the superconducting properties of the t-J model have included a comprehensive diagrammatic study of the case of low *electronic* density $\langle n_e \rangle \ll 1$ (hole doping $x\sim 1$) [160,161]. This study has shown a rich phase diagram in the $J/t - \langle n \rangle$ plane, having regions of $s$-, $p$-, and $d$-wave superconductivity and of phase separation. Later, this phase diagram was confirmed by a numerical Green's function Monte Carlo method [162]. From the RPA treatment of the Hubbard model in the strong-coupling limit, the model of "spin-bags" interacting via longitudinal magnetization fluctuations has been proposed [52] and superconductivity due to this interaction has been discussed. More recently, the transverse spin fluctuations have been considered within the same approach [163]. A diagrammatic technique for the projection operators was employed to study the t-t'-J model and the superconducting $d$-wave part in the phase diagram has been found [164].

An interesting suggestion has been put forward recently by Batyev [165,166], who considered a short-range spin-liquid state consisting of AF droplets as a possible ground state for spins in the superconducting state. A single hole in such a droplet corresponds to a spin-polaron (fermion), while two holes form a bi-polaron (boson). This allows a boson-fermion model to be derived from the t-J model. Such a model has been proposed phenomenologically by Rumer [167], without a microscopic derivation and long before the high-$T_c$ era; it was rediscovered recently by several groups [168-172]. The superconductivity arises naturally in this model as a result of the virtual transformations of two fermions into a resonance-like boson state above the Fermi sea [165].

As we discussed in the previous sections, the single- and two-hole properties of the t-J model are understood in detail within the spin-polaron paradigm. Therefore, one would seek for a more systematic approach to superconductivity in the t-J model which would make use of that understanding. A natural approach would be to integrate out the spin background and to reformulate the t-J model as an effective quasiparticle model where spin fluctuations are included in the "dressing" of the quasiparticles and in the effective hole-hole interaction. One of the first such attempts was made by Shraiman and Siggia [173,174] who used a semi-classical hydrodynamic approach to the t-J model to obtain a phenomenological Hamiltonian for the mobile vacancies coupled by the four-fermion interaction originating from the long-range dipolar twist of the spin background generated by the spin waves. The mean-field analysis of this model suggested an $s$- or $d$-wave superconducting state possibly coexisting with the incommensurate spiral AF order. Note that this work is different from the "spin-bag" approach [52] in that the interaction originates from the fluctuation of the direction, rather than the magnitude, of the magnetization.

A phenomenological model, which included the single-hole dispersion term to mimic the quasiparticle dispersion in the t-J model and a simple nearest-neighbor attraction to imitate an attractive interaction, has been studied numerically in Ref. [175]. This work has emphasized the importance of the van Hove singularity in the hole density of states for the *d*-wave superconductivity.

In another study, the hole-hole and the residual hole-magnon interactions have been obtained using an ansatz for the spin-polaron wave function, and then the effective Hamiltonian for the polarons and long-range spin waves has been presented [145]. Yet another approach used a generalization of the canonical transformation and led to the derivation of the quasiparticle Hamiltonian for interacting spin-polarons from the original t-J model [142]. This effective model describes the spin polarons and "bare" magnons interacting via polaron-polaron and a "residual" polaron-magnon interaction:

$$H_{eff} = \sum_{\mathbf{k}} \varepsilon_{\mathbf{k}} \tilde{h}_{\mathbf{k}}^{\dagger} \tilde{h}_{\mathbf{k}} + \sum_{\mathbf{q}} \omega_{\mathbf{q}} \alpha_{\mathbf{q}}^{\dagger} \alpha_{\mathbf{q}} + \sum_{\mathbf{kk'q}} V_{\mathbf{kk'q}} \tilde{h}_{\mathbf{k-q}}^{\dagger} \tilde{h}_{\mathbf{k'+q}}^{\dagger} \tilde{h}_{\mathbf{k'}} \tilde{h}_{\mathbf{k}} + \sum_{\mathbf{k}} g_{\mathbf{kq}} \left( \tilde{h}_{\mathbf{k-q}}^{\dagger} \tilde{h}_{\mathbf{k}} \alpha_{\mathbf{q}}^{\dagger} + H.c. \right), \quad (30)$$

in terms of the polaron ($\tilde{h}$) and magnon ($\alpha$) operators. It has been noted in Refs. [142] that the hole-hole interaction $V_{\mathbf{kk'q}}$ is strongly repulsive in the *s*-wave and attractive in the *d*-wave channel. Phase separation in such a model in a physical range *t>J* seems unlikely since the pair-pair interaction should also be repulsive, which agrees with the numerical data [176]. The ground state in this model would be a dilute "gas" of *d*-wave spin-polaron pairs. The BCS-type analysis of the model has been performed and the superconducting instability has been found [177]. Pairing has been obtained for *d*-wave symmetry of the gap with $T_c \sim 100$ K for realistic parameters and doping. In these works it was assumed that the antiferromagnetic order is preserved at all scales relevant to pairing. It was also demonstrated that the gap value obtained is of the order of the Fermi energy (~*xJ*). In this case, virtually all holes are involved in pairing and the integration over the **k**-space of the whole Brillouin zone is important. Therefore, the situation is quite different from the usual BCS case where $\Delta/E_F \ll 1$ is small and only the vicinity of the Fermi surface is important in **k**-space. Thus, a strong-coupling study is necessary.

Such a study has been done in the works by Plakida *et al.* [178-180] using the Eliashberg formalism applied to the t-t'-J model within the spin-polaron formulation. This approach used the original t-t'-J model and applied diagrammatic SCBA rules to the normal and anomalous Green's functions to obtain the self-consistent equations. The SCBA exploits an effective analog of the Migdal theorem for the t-J model mentioned in Sec. 4(a). The self-consistent equations are then solved numerically. The results of this study are reproduced in our Fig. 13, which shows $T_c$ versus doping for three different values of *t'* and *J=0.4t* (compare with Fig. 6 of Sec. 3).

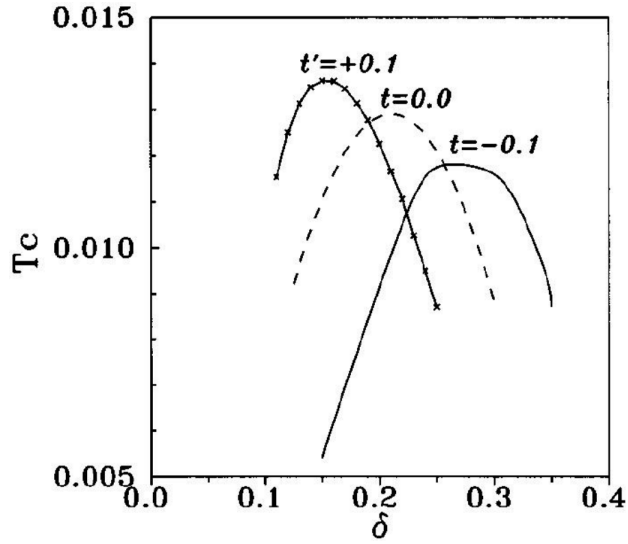

Figure 13. The superconducting temperature $T_c$ in units of $t$ versus hole concentration $\delta$ for $J=0.4t$ and $t'/t=-0.1;0;+0.1$. After Ref. [179]. Note the shift of the vertical axes from zero.

We note, however, that there exists a concern regarding these works [181]. The model used in the latter study as well as the effective models discussed above are derived assuming the long-range AF order in the system, while superconductivity occurs when the long-range AF order is lost. Nevertheless, one may argue that it is the short-range AF excitations that are most important for pairing in the spin-polaron approach. Therefore, the assumption of the long-range order is just a helpful route which allows one to apply the theoretical approach and to advance our understanding of the problem without altering the essential physics. The condition for the applicability of the above results can be rephrased as a requirement for the magnetic correlation length to be the largest length-scale in the system $\xi_{AF} \gg p_F^{-1}$, $R_p$, $\lambda_{SC}$, etc., where $p_F \propto \sqrt{x}$ is the Fermi momentum, $R_p$ is the polaron radius, and $\lambda_{SC}$ is the superconducting coherence length. Experimentally, all these lengths are of order of a few lattice spacings and it is quite possible that the required inequality is fulfilled for the real $CuO_2$ planes, as revealed by the neutron-scattering experiments [33]. However, an equivalent self-consistent *theoretical* formulation of the problem in the absence of long-range magnetic order is very complicated. Thus, a simultaneous description of the spin-liquid state of spins and of the $d$-wave superconducting state of holes remains an open problem.

#### d.) Stripes

A recent boost of interest in strongly-correlated models is due to the discovery of stripes (spin and charge inhomogeneities) in high-$T_c$ materials, where they coexist with antiferromagnetism and superconductivity [182,183]. Here we give a brief overview on how the stripes can be understood within the spin-polaron concept.

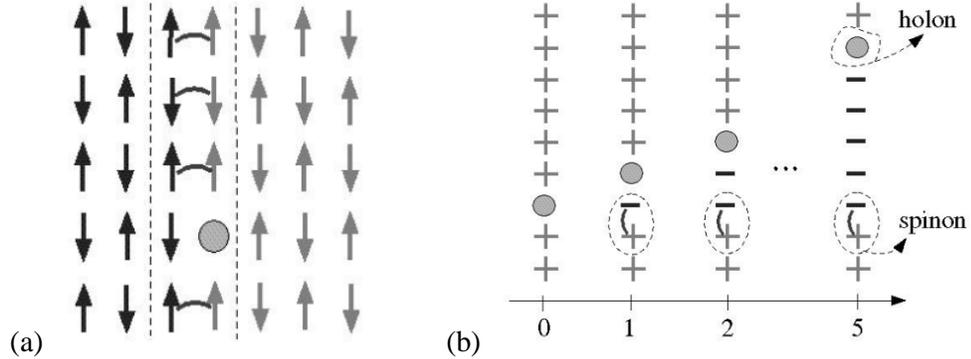

Figure 14. (a) A single hole at the anti-phase domain wall. Arcs denote "wrong" bonds. (b) Propagation of the hole in the Ising chain. Numbers indicate the number of hoppings made by the hole away from its origin, "+" and "-" represent the sign of the staggered magnetization. From Ref. [57].

In the previous subsection, we presented a generalization of the spin-polaron picture to a finite concentration of holes. Such a generalization relied on the assumption that the antiferromagnetic background remains unchanged. However, it was known from the studies of the strongly-correlated models that the "feedback" effect of holes on the antiferromagnetic background is important. Aside from Hartree-Fock treatments of the Hubbard model [184], which showed stripe-like domain wall solutions, studies of the t-J model in the low-doping regime have indicated instabilities of the antiferromagnetic order [185]. These instabilities were thought to lead towards spiral [186], stripe-like spiral [187], or spin-liquid [188] states. Earlier numerical works in the small t-J clusters, Ref. [189], have demonstrated stripes in the ground state which were also domain walls in the Néel AF. With the mounting evidence from experiments [182,183] and from Density Matrix Renormalization Group (DMRG) numerical data [190], the idea of *topological doping* [191] has flourished. The spontaneously created anti-phase domain walls have been widely considered as the topological alternatives to the homogeneous Néel background [192-195]. Thus, the many-hole ground state has turned out to be very different from the one for a few holes. In order to understand the nature of the charge excitations in the stripe phase, one needs to reconsider the single-particle problem around this new ground state with the domain wall in it [196], Fig. 14(a). The 1D character of the charge stripes has led to a number of attempts to generalize the physics of 1D systems, where the excitations are holons and spinons, as shown in Fig. 14(b), to higher dimensions [197,198]. On the other hand, there is a growing understanding that the stripes are the outcome of the same tendencies which are seen already for the single-hole problem [199], and that the charge excitations in the stripe phase may still have much in common with spin polarons [200,201].

An attempt to integrate some of the earlier ideas on the t-J model physics with the newer trends and phenomenology which have appeared due to stripes has been made recently for the t-$J_z$ model [57]. As we discussed earlier, the isotropic *SU(2)* t-J and anisotropic t-$J_z$ models lead to similar results, as long as one is concerned with the short-range physics of the charge and spin excitations. One may argue that this is also true for stripes since the stripe images obtained numerically for the t-$J_z$ model are virtually identical to the ones in the t-J model [190]. It has also been concluded, based on the Ginzburg-Landau functional approach, that the antiphase shift of the antiferromagnetic order parameter must originate from some short-range physics [202]. The rigidity of the $\pi$-shift of the antiferromagnetic phase across the domain wall in both numerical and experimental studies also argues for the short-range genesis of the stripes. In Ref. [57], the problem has been approached using a microscopic study of the stripe in an AF insulator by DMRG and by an analytical self-consistent Green's function technique developed earlier [200], which accounts for the retraceable-path motion of the holes away from stripe. The general conclusion of this study is that the stripe should be considered as a collective bound state of the holes with an antiphase domain wall. In such a system, the excitations are composite *holon-spin-polarons* which populate an effective 1D band. This picture is in very good agreement with the numerical results and provides new insight into the problem of the origin of the stripes and of the nature of electronic excitations in the stripe phase.

Briefly, the microscopic structure of the charge excitation within the stripe is given by the "longitudinal" and "transverse" components, which correspond to the motion along and perpendicular to the stripe, respectively, see Figs. 15(a,b). The longitudinal motion of the charge is equivalent to the motion of the 1D holon in the spin chain, Fig. 14(b), and thus is "free". The "transverse" motion is equivalent to the string-like motion of the hole within the spin polaron (compare Fig. 15(b) and Fig. 2). Notably, the first component in such a string is a spinon.

The coupling between these components can be expressed diagrammatically as shown in Fig. 15(c). Such a coupling leads to a renormalization of the "bare" holon Green's function [200] $G(\omega) = (\omega - 2t\cos(k_y) - \Sigma(\omega))^{-1}$ and to an effective 1D band for the low-energy holon-spin-polaronic excitations. The energy of the bottom of this band can be compared directly to the DMRG data. The results of such a comparison are shown in Fig. 16(a). The largest discrepancy between the theory and numerical data is *0.1%*, which is closer than the agreement for the single-polaron ground state presented in Sec. 4(a). This is because of the absence of the Trugman loops in the case of a stripe. Assuming a rigid-band filling of this effective 1D band, one can obtain the energy of the stripe as a function of the hole density within the stripe. This leads to the results shown in Fig. 16(b), which are also in very good agreement with numerical data. These results emphasize the primary role of the kinetic energy in favoring the stripe as a ground state.

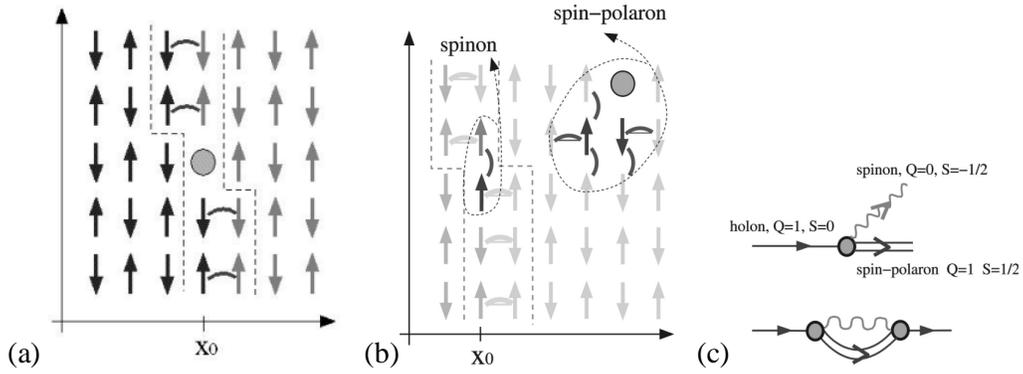

Figure 15. (a) Motion of the hole along the *y* direction is equivalent to the holon motion in Fig. 14(b). (b) A schematic result of the "transverse" hole motion, which leads to the departure from the domain wall and can be considered as a decay of a holon into a spinon and a spin-polaron. (c) Diagrams associated with such a decay and with the corresponding self-energy. From Ref. [57].

The intriguing question is whether stripes introduce a new energy scale into the problem. Naively, such a scale should be governed by the kinetic *t*-term since the hole motion is made free along the 1D stripe. However, the energy of the order $\sim JL$ is paid to "prepare" such a structure. In the continuum limit, $t \gg J$, the kinetic energy of the 1D hole motion is $\langle E_{kin} \rangle = -2t + At/L^2$, where $L$ is the length of a stripe. Magnetic energy is $\langle E_J \rangle \propto JL$, and the minimum of the energy is achieved at $L_{optimal} \propto (J/t)^{1/3}$. The corresponding minimum energy is $\langle E_{min} \rangle \cong -2t + \alpha (J^2 t)^{1/3}$. One recalls an almost identical consideration of the "retraceable-path" motion of the hole by the "strings" in the spin-polaron problem given in Sec. 2(c), which also gives $L_{string} \propto (J/t)^{1/3}$ and $\langle E_{sp} \rangle \cong -2\sqrt{3}\, t + \beta (J^2 t)^{1/3}$. Therefore, there is *no new energy scale*, different from the spin-polaron problem, introduced by the domain wall. Therefore, the "prepared-path" motion in the 1D anti-phase domain wall is, in fact, not too different from the "retraceable-path" motion in the spin polaron. One may conclude that the same scale $\propto (J^2 t)^{1/3}$ should govern the energetic balance favoring the stripe as a ground state. This is yet another argument that the stripes are the outcome of the same tendencies which are seen already for the single-hole problem.

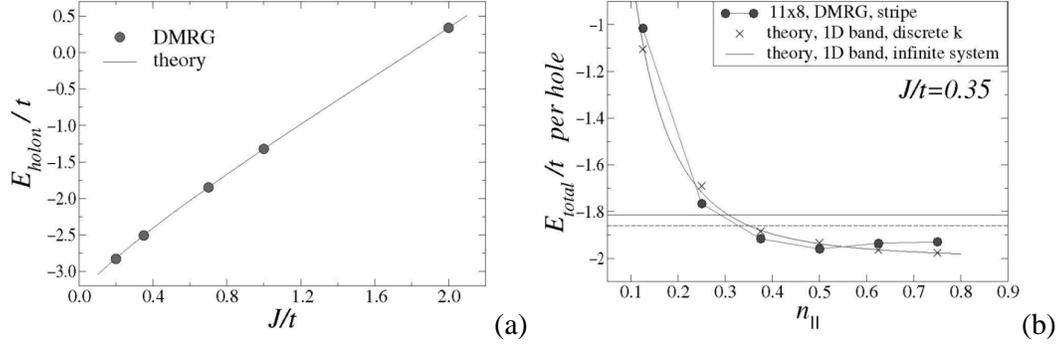

Figure 16. (a) *J/t* dependence of the single-hole ground-state energy in the stripe configuration. (b) Total energy of the system with an anti-phase domain wall per hole versus hole density. Solid curve and crosses are the theoretical results for the rigid-band filling of the 1D holon-spin-polaron band. Circles are the DMRG data from *11x8* cluster. Horizontal solid and dashed lines are the energies of free spin-polarons and bi-polarons in the homogeneous AF, respectively. From Ref. [57].

Altogether, the comprehensive comparison of the results of the theory and DMRG numerical approach presented in Ref. [57] has shown a very close quantitative agreement, thus providing strong support to this way of understanding the charge excitations at the anti-phase stripe in an antiferromagnet.

As it follows from this study, the stripe can be described by the deep "backbone" states which minimize the energy of the anti-phase configuration in the AF, and the shallow, almost free spin-polaron-like excitations around the anti-phase domain wall. Since the spin polarons are known to have a considerable pairing between themselves, such a framework does not require the superconducting pairing to come from some 1D instability, but rather suggests that the pairing is largely unrelated to the 1D stripe pattern. Such a scenario is also discussed in other recent works, Ref. [203,204]. Another advantage of this picture is a more effective screening of the long-range component of the Coulomb repulsion, which represents a problem for a system of strictly 1D charges [205].

## 5. SUMMARY AND CONCLUDING REMARKS

An overview of the spin-polaron concept as it relates to high-$T_c$ superconductivity and other phenomena in the cuprates has been given. Some features of the single-particle excitations in the $CuO_2$ planes within the framework of the multiband Hubbard and t-J models were analyzed in considerable detail. Two distinct types of spin polarons, called ferrons and string-polarons, and their similarities and differences were discussed.

Considerations of the electronic structure of the $CuO_2$ planes based on a phenomenological multiband model that assumed antiferromagnetic ordering were introduced. The dispersion curves near the top of the lower Mott-Hubbard band, renormalized by a constant effective mass ascribed to spin-polaron formation, provided good fits to ARPES data for $Sr_2CuO_2Cl_2$. Singlet pairing of two spin polarons was assumed to occur because of the local "repair" of AF ordering when the polarons were in the first and second neighbor cells of the magnetic lattice. First, the symmetry of the gap and then the dependence of the gap on dopant concentration for this illustrative model were studied.

The results for extended t-J model calculations were then summarized and the importance of the O-O hopping integral in determining certain features of the spin-polaron dispersion curves emphasized. Attention was directed to the fundamental role of spin fluctuations in determining the width of the spin-polaron bands, as found by many authors. Studies of the binding forces and the nature of the $d$-wave pairing were reviewed and the spin-wave mediated pairing mechanism was discussed for the t-J model. The importance of detailed studies of the two-body problem was noted and results for the two-hole problem within the $t-J_z$ model were presented. Investigations of $d$-wave superconductivity induced by magnetic mechanisms were surveyed and different approaches to the problem described. Special attention has been paid to those approaches that explicitly involve spin-polaronic concepts. Since a BCS-type formalism shows that essentially all holes are involved in pairing, a strong-coupling treatment of the problem is necessary. The results of such an approach produced good agreement with the phase diagram of the cuprates.

Stripe formation in the strongly-correlated models was reviewed and its relevance to spin-polaronic physics emphasized. It was shown, using recent analytical and numerical results, that stripes originate from the same spin-polaronic tendencies already seen in the string-like hole motion of the charge carrier in an antiferromagnet. The issues of the energetic balance favoring the stripes and the pairing problem in the stripe phase were addressed.

We conclude that, altogether, the spin-polaron paradigm provides a natural description of a variety of phenomena in the cuprates. Moreover, we note that the spin-polaronic results obtained from the extended Hubbard and t-J models are quite consistent with one another, as demonstrated here by the close similarities of Figs. 5 and 7 for the ARPES work and Figs. 6 and 13 for the dependence of $T_c$ on the dopant concentration. Despite the apparent success of the spin-polaron approach, it must be clear that the proper treatment of the ground state and excitation properties in the strongly correlated models of high-$T_c$ materials remains a field of extraordinarily active research, with several unresolved issues. In our view, the main open problems include:

- A self-consistent description of the $d$-wave superconducting phase coexisting with the short-range AF spin-liquid and stripes.
- The dynamics of the stripes and their relation to superconductivity.
- A consistent theory of ARPES in strongly correlated systems.

We are confident that the spin-polaron approach to high-$T_c$ superconductivity will prove of great value in resolving these and other related issues.

## ACKNOWLEDGMENTS

This research was supported by Oak Ridge National Laboratory, managed by UT-Battelle, LLC, for the U.S. DOE under contract DE-AC05-00OR22725.